\documentclass[conference,compsoc]{IEEEtran}
\usepackage[english]{babel} % handle hyphenation
\usepackage[utf8]{inputenc}
\usepackage{blindtext}

\usepackage{hyperref}

\usepackage{algorithmic}

\usepackage[ruled,vlined,linesnumbered]{algorithm2e}

\usepackage{placeins} % use \FloatBarrier to block the text float


\usepackage{amsmath}
\usepackage{amsfonts}
\usepackage{enumitem}
\usepackage{color}
\usepackage{colortbl}

\usepackage{graphicx}
\usepackage{textcomp}
\usepackage{xcolor}
\usepackage{subfigure} 
\usepackage{pifont}
\usepackage{tikz}
\usepackage{tabularray}
\usepackage{multirow}

\usepackage{amssymb}
\usepackage{soul}

\usepackage[all]{nowidow}

\def\BibTeX{{\rm B\kern-.05em{\sc i\kern-.025em b}\kern-.08em
    T\kern-.1667em\lower.7ex\hbox{E}\kern-.125emX}}

\definecolor{codepurple}{rgb}{1,0,1}
\usepackage[most]{tcolorbox}
\usepackage{tikz}

\lstdefinestyle{mystyle}{
  commentstyle=\color{codegreen},
  keywordstyle=\color{magenta},
  stringstyle=\color{codepurple},
  basicstyle=\ttfamily\scriptsize,
  breakatwhitespace=false,
  breaklines=true,
  captionpos=b,
  keepspaces=true,
  showspaces=false,
  showstringspaces=false,
  showtabs=false,
  tabsize=2
}

\newcommand{\find}[1]{
\begin{tcolorbox}[tile,size=fbox,boxsep=2mm,boxrule=0pt,top=0pt,bottom=0pt,
borderline={0.6mm}{0pt}{black!66!white},colback=black!5!white]
\em #1
\end{tcolorbox}
}

\usepackage{xcolor,colortbl}

\usepackage{xargs}
\usepackage[colorinlistoftodos,prependcaption,textsize=normalsize]{todonotes}

\newboolean{COMMENTSON} 
\setboolean{COMMENTSON}{true}   
\ifthenelse{\boolean{COMMENTSON}}
{

}

\usepackage{listings}
\usepackage[skip=1pt]{caption}

\newcommand{\tech}{\mbox{\textsc{PolyFlow}}}

\newcommand{\cvxopt}{\href{https://github.com/cvxopt/cvxopt}{Cvxopt}}
\newcommand{\mylink}[2]{\begingroup\hypersetup{pdfborder={0 0 0}}\href{#1}{\textcolor{blue}{#2}}\endgroup}
\newcommand{\Bounter}{\mylink{https://github.com/RaRe-Technologies/bounter}{Bou}}

\newcommand{\Simplejson}{\mylink{https://github.com/simplejson/simplejson}{Sim}}
\newcommand{\Japronto}{\mylink{https://github.com/squeaky-pl/japronto}{Jap}}
\newcommand{\Pygit}{\mylink{https://github.com/libgit2/pygit2}{Pyg}}

\newcommand{\Cvxopt}{\mylink{https://github.com/cvxopt/cvxopt}{Cvx}}

\newcommand{\Pytables}{\mylink{https://github.com/PyTables/PyTables}{PyT}}
\newcommand{\Pyo}{\mylink{https://github.com/belangeo/pyo}{Pyo}}

\newcommand{\Libsmbios}{\mylink{https://github.com/dell/libsmbios}{Lib}}
\newcommand{\Tink}{\mylink{https://github.com/google/tink}{Tin}}

\newcommand{\Ultrajson}{\mylink{https://github.com/ultrajson/ultrajson}{Ult}}
\newcommand{\Aubio}{\mylink{https://github.com/aubio/aubio}{Aub}}
\newcommand{\Bottleneck}{\mylink{https://github.com/pydata/bottleneck}{Bot}}
\newcommand{\Pycurl}{\mylink{https://github.com/pycurl/pycurl}{Pyc}}
\newcommand{\Msgpack}{\mylink{https://github.com/msgpack/msgpack-python}{Msg}}

\newcommand{\Jep}{\mylink{https://github.com/ninia/jep}{Jep}}

\newcommand{\Bounterall}{\mylink{https://github.com/RaRe-Technologies/bounter}{Bounter}}

\newcommand{\Simplejsonall}{\mylink{https://github.com/simplejson/simplejson}{Simplejson}}
\newcommand{\Japrontoall}{\mylink{https://github.com/squeaky-pl/japronto}{Japronto}}
\newcommand{\Pygitall}{\mylink{https://github.com/libgit2/pygit2}{Pygit2}}

\newcommand{\Cvxoptall}{\mylink{https://github.com/cvxopt/cvxopt}{Cvxopt}}

\newcommand{\Pytablesall}{\mylink{https://github.com/PyTables/PyTables}{PyTables}}
\newcommand{\Pyoall}{\mylink{https://github.com/belangeo/pyo}{Pyo}}

\newcommand{\Libsmbiosall}{\mylink{https://github.com/dell/libsmbios}{Libsmbios}}
\newcommand{\Tinkall}{\mylink{https://github.com/google/tink}{Tink}}

\newcommand{\Ultrajsonall}{\mylink{https://github.com/ultrajson/ultrajson}{Ultrajson}}
\newcommand{\Aubioall}{\mylink{https://github.com/aubio/aubio}{Aubio}}
\newcommand{\Bottleneckall}{\mylink{https://github.com/pydata/bottleneck}{Bottleneck}}
\newcommand{\Pycurlall}{\mylink{https://github.com/pycurl/pycurl}{Pycurl}}
\newcommand{\Msgpackall}{\mylink{https://github.com/msgpack/msgpack-python}{Msgpack}}

\newcommand{\Jepall}{\mylink{https://github.com/ninia/jep}{Jep}}

\newcommand{\Lwjgl}{\mylink{https://github.com/LWJGL/lwjgl}{Lwj}}
\newcommand{\Mdsplus}{\mylink{https://github.com/MDSplus/mdsplus}{Mds}}
\newcommand{\Pijv}{\mylink{https://github.com/PI4J/pi4j-v2}{Pi4}}
\newcommand{\PojavLauncher}{\mylink{https://github.com/PojavLauncherTeam/PojavLauncher}{Poj}}
\newcommand{\Turbovnc}{\mylink{https://github.com/TurboVNC/turbovnc}{Tur}}
\newcommand{\Themis}{\mylink{https://github.com/cossacklabs/Themis}{The}}

\newcommand{\Espeakng}{\mylink{https://github.com/espeak-ng/espeak-ng}{Esp}}

\newcommand{\Junixsocket}{\mylink{https://github.com/kohlschutter/junixsocket}{Jun}}
\newcommand{\Androidgifdrawable}{\mylink{https://github.com/koral--/android-gif-drawable}{And}}
\newcommand{\ImBlockerFabric}{\mylink{https://github.com/mrjesen/ImBlockerFabric}{ImB}}
\newcommand{\Hinn}{\mylink{https://github.com/switch-iot/hin2n}{Hin}}
\newcommand{\Termuxx}{\mylink{https://github.com/termux/termux-x11}{Ter}}
\newcommand{\Openjpeg}{\mylink{https://github.com/uclouvain/openjpeg}{Ope}}
\newcommand{\Wolfssljni}{\mylink{https://github.com/wolfSSL/wolfssljni}{wol}}

\newcommand{\Lwjglall}{\mylink{https://github.com/LWJGL/lwjgl}{Lwjgl}}
\newcommand{\Mdsplusall}{\mylink{https://github.com/MDSplus/mdsplus}{Mdsplus}}
\newcommand{\Pijvall}{\mylink{https://github.com/PI4J/pi4j-v2}{Pi4j-v2}}
\newcommand{\PojavLauncherall}{\mylink{https://github.com/PojavLauncherTeam/PojavLauncher}{PojavLauncher}}
\newcommand{\Turbovncall}{\mylink{https://github.com/TurboVNC/turbovnc}{Turbovnc}}
\newcommand{\Themisall}{\mylink{https://github.com/cossacklabs/Themis}{Themis}}

\newcommand{\Espeakngall}{\mylink{https://github.com/espeak-ng/espeak-ng}{Espeak-ng}}

\newcommand{\Junixsocketall}{\mylink{https://github.com/kohlschutter/junixsocket}{Junixsocket}}
\newcommand{\Androidgifdrawableall}{\mylink{https://github.com/koral--/android-gif-drawable}{Android-gif-drawable}}
\newcommand{\ImBlockerFabricall}{\mylink{https://github.com/mrjesen/ImBlockerFabric}{ImBlockerFabric}}
\newcommand{\Hinnall}{\mylink{https://github.com/switch-iot/hin2n}{Hin2n}}
\newcommand{\Termuxxall}{\mylink{https://github.com/termux/termux-x11}{Termux-x11}}
\newcommand{\Openjpegall}{\mylink{https://github.com/uclouvain/openjpeg}{Openjpeg}}
\newcommand{\Wolfssljniall}{\mylink{https://github.com/wolfSSL/wolfssljni}{wolfssljni}}

\definecolor{delcolor}{rgb}{1.0, 0.8, 0.8}
\definecolor{addcolor}{rgb}{0.8, 1.0, 0.8}
\definecolor{deltext}{rgb}{0.86, 0.08, 0.24}
\definecolor{addtext}{rgb}{0, 0.5, 0}

\lstdefinelanguage{diff}{
  morecomment=[f][\color{blue}]{@@},
  morecomment=[f][\color{deltext}]{-},
  morecomment=[f][\color{addtext}]{+},
  morecomment=[f][\color{deltext}]{---},
  morecomment=[f][\color{addtext}]{+++},
}

\lstdefinestyle{mystyle}{
    basicstyle=\ttfamily\scriptsize,
    columns=fullflexible,
    backgroundcolor=\color{white},
    numbers=none,
    showstringspaces=false,
    escapeinside={(*@}{@*)},
    frame=none,
    keywordstyle=\color{blue},
    moredelim=[is][\color{deltext}\bfseries\colorbox{delcolor}]{\%-}{-\%},
    moredelim=[is][\color{addtext}\bfseries\colorbox{addcolor}]{\%+}{+\%},
    language=diff,
    aboveskip=0pt,
    belowskip=0pt,
    aboveskip=-3pt,  % Adjust as needed
    belowskip=-3pt,   % Adjust as needed
}

\definecolor{darkgreen}{rgb}{0,0.5,0}
\definecolor{darkgray}{rgb}{0.4,0.4,0.4}
\definecolor{darkred}{rgb}{0.6,0,0}

\usepackage{xspace}   
\usepackage{graphicx}
\usepackage{booktabs} % To thicken table line
\usepackage{multirow}

\usepackage{wrapfig}
\usepackage{diagbox}
\usepackage{algorithmic}
\usepackage{graphicx}
\usepackage{listings}
\usepackage{xcolor}
\usepackage{tikz}
\usepackage[framemethod=tikz]{mdframed}
\usepackage[most]{tcolorbox}
\usepackage{pifont}
\usepackage{enumitem}
\usepackage{wrapfig}
\usepackage{url}
\usepackage{amsfonts}
\usepackage{stmaryrd}
\usepackage{tikz}
\usepackage[strings]{underscore}

\definecolor{blush}{rgb}{0.87, 0.36, 0.51}
\newcommand{\etal}{{et al.}\xspace}

\usepackage{colortbl}

\definecolor{codegreen}{rgb}{0,0.6,0}
\definecolor{codegray}{rgb}{0.5,0.5,0.5}
\definecolor{codepurple}{rgb}{0.58,0,0.82}
\definecolor{backcolour}{rgb}{0.95,0.95,0.92}

\definecolor{winered}{rgb}{0.7,0,0}
\definecolor{gray}{gray}{0.7}
\definecolor{darkpastelgreen}{rgb}{0.01, 0.75, 0.24}
\definecolor{cadmiumgreen}{rgb}{0.0, 0.42, 0.24}
\definecolor{brickred}{rgb}{0.8, 0.25, 0.33}
\definecolor{cornellred}{rgb}{0.7, 0.11, 0.11}
\definecolor{burgundy}{rgb}{0.5, 0.0, 0.13}
\definecolor{frenchblue}{rgb}{0.0, 0.45, 0.73}
\definecolor{light-gray}{gray}{0.92}
\definecolor{lightlight-gray}{gray}{0.97}
\definecolor{codegray}{gray}{0.90}
\definecolor{inputgray}{gray}{0.90}
\definecolor{darkgreen}{RGB}{40,125,40}
\definecolor{delim}{RGB}{20,105,176}
\definecolor{blizzardblue}{rgb}{0.67, 0.9, 0.93}

\definecolor{amethyst}{rgb}{0.5, 0.3, 0.7}

\lstdefinestyle{mystyle}{
  backgroundcolor=\color{white},   commentstyle=\color{codegreen},
  xleftmargin=.03\textwidth,
  xrightmargin=.02\textwidth,
  keywordstyle=\color{magenta},
  numberstyle=\tiny\color{darkgray},
  stringstyle=\color{codepurple},
  basicstyle=\ttfamily\footnotesize,
  breakatwhitespace=false,         
  breaklines=true,                 
  captionpos=b,                    
  keepspaces=true,                 
  numbers=left,                    
  numbersep=5pt,                  
  showspaces=false,                
  showstringspaces=false,
  showtabs=false,                  
  tabsize=2
}

\definecolor{dkgreen}{rgb}{0,0.6,0}
\definecolor{gray}{rgb}{0.5,0.5,0.5}
\definecolor{mauve}{rgb}{0.58,0,0.82}

\global\mdfdefinestyle{rtboxstyle}{%
linecolor=black,%
leftmargin=0cm,rightmargin=0cm,linewidth=0.5pt,
roundcorner=3,
skipbelow=0pt,backgroundcolor=lightlight-gray
}

\makeatletter
\newcommand\figcaption{\def\@captype{figure}\caption} 
\newcommand\tabcaption{\def\@captype{table}\caption} 
\makeatother

\newcommand{\yes}{\tikz\draw[fill=black] (0,0) circle (0.6ex);}
\newcommand{\no}{\tikz\draw[draw=black,fill=white] (0,0) circle (0.6ex);}
\newcommand{\rhalf}{\tikz{
    \draw[draw=black,fill=white] (0,0) circle (0.6ex);
    \clip (-0.6ex,-0.6ex) rectangle (0,0.6ex); % left half fill
    \fill[black] (0,0) circle (0.6ex);}
}

\usepackage{cleveref}

\usepackage{diagbox}
\usepackage{pifont}
\newcommand{\cmark}{\checkmark}

\usepackage[font=scriptsize]{subcaption}

\usepackage[
  frozencache=true,
  cachedir=minted-cache
]{minted}
\newtcbox{\codebox}{on line, boxrule=0.1pt, top=0pt,bottom=0pt, left=0pt, right=0pt, arc=1pt, fontupper=\scriptsize}
\setminted{fontsize=\scriptsize,baselinestretch=1,frame=lines,framesep=2mm}

\definecolor{codehighlight}{RGB}{255,255,200}

\begin{document}

\author{\IEEEauthorblockN{Haoran Yang}
\IEEEauthorblockA{Washington State University
%\\yhryyq@gmail.com
}
\and
\IEEEauthorblockN{Zhixuan Zhong}
\IEEEauthorblockA{University at Buffalo, SUNY
%\\mc.b0x.too.f00l@gmail.com
}
\and
\IEEEauthorblockN{Jiawei Guo}
\IEEEauthorblockA{University at Buffalo, SUNY
%\\jiaweigu@buffalo.edu
}
\and
\IEEEauthorblockN{Haipeng Cai%{$^\text{\Letter}$}
}
\IEEEauthorblockA{University at Buffalo, SUNY
%\\haipengc@buffalo.edu}
}}

%\title{{\tech}: Agent-Based Static Information Flow Analysis Across Language Boundaries}
%\title{{\tech}: An Agentic Framework for Static Information Flow Analysis across Language Boundaries}
\title{%{\tech}: 
{\mbox{\textsc{PolyFlow}}}: 
A Neuro-Symbolic Framework for Static Cross-Language Information Flow Analysis}

\maketitle

\begin{abstract}

Modern software systems are commonly constructed in multiple, interacting programming languages. %, often connected via foreign function interfaces such as Python extensions and Java JNI bindings. 
This construction leads to additional, often stealthy vulnerabilities buried in complex information flow due to language interactions. Existing static analyzers are impeded by the heterogeneous semantics of different languages, whereas dynamic approaches suffer from the limited coverage of (available and/or generated) test inputs. In this paper, we develop {\tech}, a neural-symbolic framework for statically reasoning about information flow across language boundaries, combining large language models (LLMs) and static analysis synergistically. Governed by the control-flow representation of a given multi-language system, {\tech} leverages LLMs to identify implicit flow facts due to challenging language features, hence augmenting the base representation and then propagating data flow through the system. It tackles inherent barriers (e.g., token limit and hallucination) of LLMs by putting them under careful guidance (e.g., static-analysis-guided scoping, context management, and fact checking), along with a multi-LLM expert panel for negotiated validation. 
%We instantiated \tech on both Python–C and Java–C systems to evaluate its multi-language analysis generality. 
%
%We implemented {\tech} for Python–C and Java–C systems. 
Our experiments %across these two impactful language combinations 
on real-world Python–C and Java–C systems
show that {\tech} is cost-effective and superior to various kinds of state-of-the-art baselines, revealing previously unknown cross-language vulnerabilities %in these systems 
that are missed by all the baselines.
\end{abstract}

\thispagestyle{plain}
\pagestyle{plain}
\pagenumbering{arabic}
% Use the following at camera-ready time to suppress page numbers.
% Comment it out when you first submit the paper for review.
%\thispagestyle{empty}
%\pagestyle{empty}

% \IEEEpeerreviewmaketitle

%!TEX root = paper.tex
%\vspace{-3pt}
\section{Introduction}\label{sec:intro}
%\vspace{-2pt}
% the context of problem: prevalance of multilingual construction
Software systems today are mostly and increasingly \textit{multilingual}~\cite{jones2010software,delorey2007programming,ray2014large,yang2024multi}, composed of code units %written 
in diverse programming languages that \textit{interact} with each other to deliver seamless functionalities~\cite{tomassetti2014empirical,mayer2015empirical,valverde2015punctuated,li2024multilingual,haoran23icse,yangtse24}. This heterogeneous composition is well justified by combining the unique advantages of individual languages\cite{bae2019towards,mayer2017multi}. 
For instance, highly impactful machine learning frameworks (e.g., PyTorch) and mobile systems (e.g., Android) %(e.g., TensorFlow and PyTorch) 
are typical examples of multi-language software that couple a high-level host language
(e.g., Python or Java) with performance-critical native C code via foreign-function interfaces~\cite{hu2023empirical,li2023understandingb,li2025automated}. 
%Such systems leverage high-level languages for rapid development and rich ecosystems, while relying on C for efficiency and low-level control.\cite{hu2023empirical,li2023understandingb,li2025automated}. 

% the problem targeted: presence and prevalence and damage of cross-language vulnerabilities
However, %so much as it enhances productivity, 
the multilingual construction also results in greater complexity of multi-language systems, 
which leads to their statistical proneness~\cite{wen22fse,ray2014large,abidi2021multi,kochhar2016large,grichi2020impactjni} and practical exposure~\cite{dinh2021favocado,wen22usenixsecurity} to increased security threats.
In particular, the interaction between languages introduces \textit{{cross-language vulnerabilities}} as recently discovered~\cite{staicu2023bilingual,wen23usenixsecurity}: security flaws that arise from semantic mismatches, implicit assumptions and flows, or unsafe data exchange across language boundaries~\cite{hwang2021justgen,yang2024learning}. 
For example, in Python-C systems, improper marshaling of data types due to dynamic typing in Python caused unsafe memory allocation hence corruption in C through a native function call~\cite{wen22usenixsecurity}. 
%As another example, unchecked string formatting in C led to code injection vulnerabilities in Python via a foreign function return~\cite{wen22fse,haoran25fse}.  
These vulnerabilities are systemic yet stealthy, often evading detection 
%due to the fragmented nature of existing analysis tools~\cite{horwitz10jan,yang2024learning}, which operate in language-specific silos. 
as existing analysis tools mostly operate in language-specific silos~\cite{horwitz10jan,yang2024learning}. 
Securing multi-language systems demands a %paradigm shift toward 
\textit{holistic}, cross-language analysis~\cite{wen22usenixsecurity,youn2023declarative,dinh2021favocado,wen22fsetool}, whose urgency is driven by the growing dominance of those systems~\cite{yang2024multi,meyerovich2013empirical,vasilescu2013babel,li2023understandinga,li2024exploratory}. 

In response, %the last few years have seen a number of 
relevant approaches exist. 
Yet many of them~\cite{fin2001amleto,fragoso2020gillian,groce2018extensible,liu2020fans,dinh2021favocado,moller2024cross} actually tackle single-language bugs, where the
“cross-language" terms, when present, refer to “supporting different languages", \textit{not} concerning scenarios where those languages \textit{co-exist and interact} in one project~\cite{li2023understandinga,li2023understandingb,li2024exploratory,ray2014large,berger2019impact,zhang2019study}. 
Among techniques that do address multilingual code, some only look at the design-level smells~\cite{abidi2021multi,li2025automated} or are limited to %cross-language 
API misuses~\cite{sultana2016understanding,hu2023empirical}. 
To really tackle cross-language bugs, a few \textit{dynamic} approaches emerged~\cite{bai2018bridgetaint,xue2018ndroid,wen22usenixsecurity,jiang2025powerpoly,haoran22fsenier}. However, they can only reveal bugs exercised by given test cases---which are particularly scarce in multi-language projects~\cite{wen23usenixsecurity}. 
While this may be mitigated by generating more tests, existing solutions %either 
resort to much manual effort~\cite{hwang2021justgen} or fail due to the technique's randomness~\cite{wen23usenixsecurity}. 

Current \textit{static} approaches bypass the coverage limitation, but they rely on language-specific semantic rules~\cite{youn2023declarative} or abstract summaries~\cite{lee2020broadening,park2023static}/specification~\cite{kan2024cross} for native code and/or language interoperability, hence requiring labor-intensive per-language engineering. 
Recent advances %provide supporting capabilities for 
support static cross-language analysis~\cite{roth2024axa,zhang2025interactive}, whose potential for detecting cross-language bugs has yet to be demonstrated. More importantly, they require separate, mature analysis infrastructure for each language, which is not always available (e.g., for Python). 
%Moreover, as a common barrier to static analysis in general, these approaches suffer both imprecision~\cite{wei2018jn} due to their conservative nature and limited recall due to their inability to analyze dynamic language constructs which prevalent in modern languages~\cite{wen22usenixsecurity}. 
As showcased lately, \textit{learning-based} cross-language bug detection~\cite{yang2024learning,li2025fine} may potentially complement both static and dynamic approaches. 
Unfortunately, they struggle to generalize to new and unseen codebases, and their performance is highly dependent on the quality and quantity of training data, which (e.g., labeled buggy datasets~\cite{yang2025dissecting}) are particularly lacking in the domain of multi-language software.
Additionally, extant static, dynamic, and learning-based techniques are commonly limited in analyzing challenging language features (e.g., decorators in Python, as well as reflection in Java) hence 
missing highly stealthy cross-language bugs (e.g., those buried in implicit information flow).

In this paper, we aim to \textit{statically} analyze \textit{cross-language information flow} while overcoming the %aforementioned 
limitations of extant approaches. 
Static information flow analysis 
is a fundamental technique for vulnerability discovery~\cite{myers1999jflow,ferraiuolo2017verification,jovanovic2006pixy,huang2015supor,nan2015uipicker,li2025iris}, which formulates the detection of various kinds of vulnerabilities as a source-sink problem (with sources/sinks defined accordingly). It has %drawn great attention and 
been widely used in practice (e.g., underlying GitHub CodeQL~\cite{github_codeql} and Facebook Infer~\cite{calcagno2011infer}) \textit{for single-language software}. 
Yet, %in light of the known shortcomings of each and given the complexity of multi-language code, fulfilling our aim comes with  
developing a static information flow analysis \textit{across language boundaries} 
%faces several key challenges. 
is challenging. 

First (\textit{Challenge-1}), 
different languages define and handle constructs (e.g., type systems, memory management, and control flow) differently. 
These %differences 
semantics disparities 
complicate the creation of a unified (especially static) analysis framework capable of 
accurately interpreting and reasoning about code across languages. 
Second (\textit{Challenge-2}), diverse language features (e.g., function pointers and inline assembly in C), especially %those due to 
dynamic code constructs (e.g., dynamic typing and reflection in Python, dynamic class loading in Java), 
%result in intricacies that classical static analysis techniques struggle to address, even for single-language code. 
%These features 
can obscure the flow of information and make it difficult to analyze %program behavior at compile time, even for single-language code. 
even single-language code at compile time. 
Dealing with such features in/across multiple languages %, especially those inducing complex cross-language behaviors, 
is only harder. 

These intrinsic static-analysis challenges could be tackled by large language models (LLMs), given their demonstrated potential for code understanding~\cite{fang2024large,linlarge25ndss} and reasoning~\cite{li2024enhancing,xie2024resym,li2025iris}. 
Yet (\textit{Challenge-3}), LLMs are not magic boxes---merely delegating a complex task like cross-language analysis to them may not help much. 
For one thing, their token/context-size limits %, it is not always feasible to even pass a single function to it
make it infeasible to feed them with the entire system, especially as multi-language codebases tend to be larger  
than single-language ones. Analyzing one function at a time alleviates this barrier, but the LLMs would not capture even %critical 
interprocedural context, not to mention 
inter-language effects---both of which are essential for effective cross-language analysis. 
%intraprocedural analyses 
For another, 
LLMs %are well-known to suffer hallucinations, l
may hallucinate, %likely producing unstable and even spurious results. 
producing spurious results~\cite{li2024enhancing,xie2024resym}.

Overcoming
such challenges, we developed {\tech}, \textit{a repository/project-level %context-, flow-, and path-sensitive 
static cross-language information flow analysis} that synergistically combines traditional program analysis with foundation LLMs in a neural-symbolic framework. 
%The intuition/rationale is that LLMs have shown great potential for understanding code across different languages~\cite{fang2024large,linlarge25ndss} and traditional program analysis may be effectively assisted by LLMs to deal with its inherent barriers~\cite{li2024enhancing,xie2024resym}. 
Given a multi-language system, we start with a lightweight cross-language interprocedural analysis, resulting in an approximate (incomplete and imprecise) representation of the whole system with only explicit control/data flow facts. 
Based on this representation, the static analysis works further as the system controller, invoking LLMs to analyze implicit flow facts, hence potentially augmenting the representation and propagating information between given sources and sinks through the system, so as to address \textit{Challenge-1}. 

To identify and resolve the implicit flows, which are induced by respective language features, 
we elicit the LLMs' relevant knowledge via few-shot in-context learning, while aiming to achieve analysis soundness via a fixed-point iteration of feature handling, hence addressing \textit{Challenge-2}. 

Importantly, the static analysis coordinates interprocedural and inter-language contexts, which allows the LLMs to only provide complementary assistance (e.g., 
handling the challenging features) in a minimal scope (a function or a control-flow path) on demand. 
Moreover, we leverage multiple LLMs formed as a negotiation/arbitration component, called \textit{expert panel}, to cross-validate results of every LLM query, 
while using static-analysis-produced facts to check those results. 
The careful guidance (adaptive scoping, context management, and fact checking) and negotiated-agreement-based validation procedure collectively address \textit{Challenge-3}.

For evaluation purposes, we implemented {\tech} for Python-C and Java-C systems %using a query-based static analyzer, 
given their high and lasting popularity (e.g., in the impactful AI/ML, mobile, and web software ecosystem)~\cite{wen22fse,li2024multilingual,yang2024multi}. 
To rigorously validate {\tech}, we curated \textit{xFlowBench}, the first static cross-language analysis benchmark with ground-truth information flow facts with respect to 
different features of three languages. 
We then applied {\tech} to 29 real-world multi-language systems of diverse scales and application domains, and compared it to various kinds of state-of-the-art (SOTA) 
%code-analysis- and learning-based 
baselines. %(MultiQL~\cite{youn2023declarative}, xLoc~\cite{yang2024learning}, PolyCruise~\cite{wen22usenixsecurity}, and PolyFuzz~\cite{wen23usenixsecurity}). 

Our extensive experiments reveal {\tech}'s promising cost-effectiveness and scalability. 
On the microbench, {\tech} achieved 97.9\% precision, 87.3\% recall, and 91.8\% F1. On the complex real-world systems, it achieved 76.8\% precision and discovered 17 previously unknown, exploitable cross-language vulnerabilities---none of which are found by any of the baselines. 
%of which ?? could not be found by any of the baselines
%
% ---with ?\%--?\% higher precision. 
%For source/sink identification, it attained ?\% precision and ?\% recall. 
Ablation studies confirmed that each of its key design elements/components contributes significantly to its overall performance. 
{\tech} also reasonably scales to large-scale systems, incurring 12.86 hours and 482.5 MB peak memory on average per project; 
%in comparison, the best-performing (for vulnerability discovery) baseline took ?? hours with ?? MB peak memory. 
In terms of LLM costs, {\tech} only consumes on average 2.81-M tokens 
% and \$?? 
to analyze each multi-language system. 
%
%PolyFlow achieves high precision and recall in detecting cross-language vulnerabilities, surpassing existing static and dynamic analysis techniques. PolyFlow's ability to effectively address the challenges of language semantics disparity, language diversity, dynamic language features, and cost-effectiveness makes it a practical and scalable solution for securing multilingual software systems. These results underscore the potential of LLM-guided static analysis as a powerful tool for vulnerability detection and software assurance.

%PolyFlow reduces false positives by 40\% compared to MLQL and identifies 3× more true positives than AXA. Its LLM-agent resolved 85\% of dynamic feature-induced ambiguities (e.g., implicit flows via Python’s @property decorators), validating its iterative refinement process. The hybrid approach also scales to 500k-LoC codebases, demonstrating practical applicability.

%Through {\tech}, we demonstrated
{\tech} showcases
a novel paradigm of neuro-symbolic approaches % of combining traditional static analysis with LLMs for 
%to practical cross-language information flow analysis, where unlike in typical LLM-based agents the static analysis makes main decisions (not the LLMs). 
to %cross-language 
code analysis.  
Instead of an LLM-based core invoking external (symbolic) %-reasoning) 
tools, our framework is \textit{centered around a static analyzer},
which judiciously (1) utilizes LLMs where they offer the most value while traditional analysis falls short, 
(2) decides when, where, and how to employ LLMs while supplying contexts on demand, and (3) integrates LLM outputs as intermediate results to the analysis, 
hence combining the strengths of both methods while overcoming each other's shortcomings---the \textit{key insight and takeaway} of our work. 
%
%While for design validation we only implemented it for Python-C so far, the methodology can be applied to other language combinations, especially given the support of both the underlying static analysis framework and LLMs for different languages. 
%In summary, 

The main contributions of this paper include: 

\begin{itemize}[leftmargin=*,topsep=2pt, noitemsep]
    \item {\tech}, a novel static %, fine-grained (statement-level) 
    %information flow 
    analysis of multi-language software that %synergistically integrates foundational LLMs with traditional static analysis to 
    holistically addresses cross-language %vulnerabilities, particularly those via implicit flows or 
    information flow, including implicit ones and those 
    induced by language features challenging to existing analyses ($\S$\ref{sec:technique}).

    \item An open-source implementation of {\tech} working with real-world Python–C and Java–C systems across various scales and domains ($\S$\ref{sec:implementation}).
    
    \item \textit{xFlowBench}, an open multilingual static analysis benchmark, which %to the best of our knowledge also 
    is the first such benchmark with ground-truth cross-language information flow facts %analysis techniques, which is the first multilingual static analysis benchmark suite publicly available as we know 
    as we know
    ($\S$\ref{sec:microbench}).

    \item A comprehensive evaluation of {\tech} on both the micro-benchmarks and real-world multi-language systems, which demonstrates its promising merits in  
    cross-language analysis effectiveness and vulnerability discovery superior to SOTA baselines ($\S$\ref{sec:eval}). 
\end{itemize}

\section{Background and Motivation}\label{sec:bkgmotive}
We introduce concepts in cross-language code analysis, and motivate our work via an 
example and a broader study. 
%a study that reveals challenges to static information flow analysis of multilingual code. 

\subsection{Multilingual Code Construction}

%Multi-language software refers to systems that incorporate multiple computer languages \textit{that interact}. The interactions %between languages 
%typically rely on \textit{Foreign Function Interfaces (FFIs)}, which enable cross-language communication through specific APIs. For example, Python and C each provide distinct FFI mechanisms---Python through the \textit{ctypes} library and C-extensions.

Multi-language systems are software written in multiple computer languages \textit{that interact}, where the program is referred to as \textit{multilingual code/program}.
The interactions %between languages 
typically rely on \textit{Foreign Function Interfaces (FFIs)}, a mechanism that enables code written in one (host) language to call functions written in another (guest) language. 

For example, Python provides FFIs for cross-language interfacing with C via modules like \textit{ctypes} and C-extensions. 
In particular, such interfacing involves two main types of \textit{cross-language functions}: native functions and foreign functions. Native functions are implemented by developers in a guest language to be invoked in the host language code, %different from the caller’s language, 
while foreign functions are provided by the host language to facilitate interoperability with the guest language code.

\subsection{Motivating Example}
Different from those in a single-language program, vulnerabilities in multilingual code  may be rooted in vulnerable cross-language information flow~\cite{wen22fse,staicu2023bilingual}, where the information source lies in (the code written in) one language while the sink is located in another~\cite{wen22usenixsecurity}. 
Apparently, the vulnerabilities will be missed if we only analyze one language component. 
Moreover, certain language features may lead a static analyzer to miss even the intra-language flow segments, hence breaking the entire cross-language flow.

% \begin{tcolorbox}[title=Motivating Example (from the Cvxopt project), 
%                   fonttitle=\bfseries\small, 
%                   colback=blue!5!white, 
%                   colframe=blue!50!black]
% \begin{minted}[fontsize=\scriptsize]{python}
% # Python: @app.route decorator obscures control flow
% @app.route('/process')  # SOURCE (user input)
% def process_data():
%     user_input = request.args.get('data')
%     c_module.validate(user_input) # native call
% \end{minted}

% \begin{minted}[fontsize=\scriptsize]{c}
% // C: Function pointer prevents control flow resolution
% void (*validator)(char*);  // Flow broken here

% void validate(char* input) {
%     if(validator) 
%         validator(input); // SINK (format-string vuln)
% }
% \end{minted}
% \end{tcolorbox}

\begin{figure}[t]
\centering
\begin{minipage}{0.48\textwidth}
%\begin{compactcode}
\begin{minted}[fontsize=\scriptsize]{python}
# Python: @app.route decorator obscures control flow
@app.route('/process')  # SOURCE (user input)
def process_data():
    user_input = request.args.get('data')
    c_module.validate(user_input) # native call
\end{minted}
%\caption{Python entry point with decorator}
%\label{fig:python-code}
%\end{compactcode}
\end{minipage}
\hfill
\begin{minipage}{0.48\textwidth}
%\begin{compactcode}
\begin{minted}[fontsize=\scriptsize]{c}
// C: Function pointer prevents control flow resolution
void (*validator)(char*);  // Flow broken here

void validate(char* input) {
    if(validator) 
        validator(input); // SINK (format-string vuln)
}
\end{minted}
%\caption{C validation with function pointer}
%\label{fig:c-code}
%\end{compactcode}
\end{minipage}
%\caption{An example cross-language vulnerability.}
\caption{Code snippet adapted from the {\cvxopt} project.}
\label{fig:vuln-example}
\end{figure}

Consider the code snippet in Figure~\ref{fig:vuln-example}, which is from the \texttt{Cvxopt} project. 
When handling optimization requests, user-submitted data enters through a Python endpoint registered via Flask's \colorbox{lightgray}{@app.route} 
%@app.route 
decorator, a language feature that obscures control flow, where the input is collected without proper sanitization. 
This tainted data then crosses into C through a native module interface %(native\_module.validate()). 
(\colorbox{lightgray}{native\_module.validate()}). 
Within the C validation logic, a function pointer indirection %(void (*validator)(char*)) 
(\colorbox{lightgray}{void (*validator)(char*)})     
dynamically dispatches the input to a {\tt printf}-like sink, creating a format-string vulnerability exploitable through malicious payloads. 

Existing static analyzers may miss this critical flow: Python-focused tools may fail to recognize the decorator-registered route as an entry point, while C-focused analyzers stop tracking flow at the unresolved function pointer call. Moreover, without analyzing the cross-language flow from Python's HTTP handler to C's format-string sink, this
multilingual code vulnerability will not be detected. 

% \begin{figure}[h]
%     \centering
%     \includegraphics[width=1\columnwidth]{Figures/func_pointer.pdf}
%       \caption{Function pointer real-world usage. The call path of the function pointer \code{word2bool} is marked with green lines.}
%     \label{fig:motivatingexample}
% \end{figure}
% The use of function pointers introduces indirect control-flow edges. The function \texttt{word2bool} (declared in line 1) is a utility function converting data from type \texttt{word} to \texttt{bool}, crucial for conditional decisions. It passes to the function \texttt{findletters} (line 2), which is declared in line 3, accepting function pointer \texttt{word2bool} as the third parameter \texttt{tobool} and invoking function \texttt{getsig} (line 7) declared in line 10, inside which the function \texttt{word2bool} is eventually invoked in line 19. Traditional static analyses struggle to resolve such indirect calls due to the dynamic nature of function pointers, causing incomplete CICFGs that miss these critical edges.

% start with CFG by Joern, enabled by, we define interprocedual control flow graph. based on ICFG for each language, we then connect the   for foreign language and native and get holistic graph. Each node is a statement, each edge represents a control flow relationship.

\subsection{Motivating Study}
%To systematically identify language features that challenge static analysis in cross-language environments beyond function pointers, we conducted a comprehensive investigation through three approaches: (1) analysis of prominent open-source Python-C projects, (2) review of developer discussions and technical blogs, and (3) empirical evaluation of static analysis tools. We analyzed prominent open-source Python-C projects mined from GitHub, building upon the dataset curated in the prior study on cross-language bugs~\cite{yang2025dissecting}. This analysis involved forward and backward traversals of cross-language information flows, using CICFG-based inspection to pinpoint critical locations and root causes of cross-language bugs. The study highlighted specific Python and C constructs—such as dynamic typing and language-specific APIs—that frequently hinder static analysis. This process revealed 14 problematic features across Python and C that significantly impede static analysis. We evaluated these features' impact on CFG generation by assessing Joern's ability to correctly process each case, with detailed results presented in Table \ref{tab:motivation-study}.
%
In Figure~\ref{fig:vuln-example}, the challenging features (decorators and function pointers) obstruct intra-language analysis, hence breaking the flow across the language boundary. These features often cause implicit information flow, which may bear stealthy vulnerabilities. 
To understand this problem, we intend to systematically identify such language features that challenge static analysis. %in multi-language environments. 
%We again started for Python-C, and 
We took a three-pronged study methodology: (1) analysis of prominent open-source Python-C and Java-C projects on GitHub (including those in cross-language bug datasets~\cite{wen22fse,yang2025dissecting}), (2) review of developer discussions and official language documentation, and (3) empirical evaluation of popular static analysis tools. Via manual inspection, we track cross-language information flows to pinpoint critical locations and root causes of cross-language bugs in relation to specific Python, Java and C %constructs 
features %---such as dynamic typing and language-specific APIs---
that frequently hinder the static analyzers (e.g., by breaking control/data flow tracking). 

%This process identified 14 problematic features across both languages that significantly impede static analysis, whose impact on CFG generation we evaluated through Joern's ability to correctly process each case, with detailed results presented in Table~\ref{tab:motivation-study}.

\setlength{\tabcolsep}{2pt}
\begin{table}[tp]
\centering
\caption{Challenging language features, impact on analysis, and support by various analyzers 
($\protect\yes$: support, $\protect\no$: no support, $\protect\rhalf$: partial/limited support, $\protect\cmark$: missing, -: no impact)}
\resizebox{\columnwidth}{!}{%
\renewcommand{\arraystretch}{0.8}
\scriptsize
\begin{tabular}{@{}llcccccccc@{}}
\toprule
\multirow{2}{*}{\textbf{Lang.}} & \multirow{2}{*}{\textbf{Feature}} & \multicolumn{2}{c}{\textbf{Impact}} & \multicolumn{5}{c}{\textbf{Static Analyzer Support}} & \multirow{2}{*}{\textbf{{\tech}}} \\
\cmidrule(lr){3-4} \cmidrule(lr){5-9}
& & \textbf{Ent.} & \textbf{Flow} & \textbf{Joern} & \textbf{CodeQL} & \textbf{Infer} & \textbf{SonarQube} & \textbf{Clang} \\
\midrule
\multirow{6}{*}{\rotatebox{90}{\textbf{Python}}} 
& Decorators & \cmark & \cmark & \no & \rhalf & \text{N/A} & \rhalf & \text{N/A} & \yes \\  
& First-Class Functions & -- & \cmark & \no & \rhalf & \text{N/A} & \rhalf & \text{N/A} & \yes \\  
& Lambda Functions & \cmark & \cmark & \no & \rhalf & \text{N/A} & \rhalf & \text{N/A} & \yes \\  
& Dynamic Typing & -- & \cmark & \no & \rhalf & \text{N/A} & \rhalf & \text{N/A} & \yes \\
& Reflection & \cmark & \cmark & \no & \rhalf & \text{N/A} & \no & \text{N/A} & \yes \\  
& Dynamic Imports & \cmark & \cmark & \no & \rhalf & \text{N/A} & \no & \text{N/A} & \yes \\  
\midrule

\multirow{5}{*}{\rotatebox{90}{\textbf{Java}}}
& Reflection & \cmark & \cmark & \no & \rhalf & \no & \rhalf & \text{N/A} & \yes \\
& Lambda Expressions & \cmark & \cmark & \rhalf & \yes & \rhalf & \yes & \text{N/A} & \yes \\
& Dynamic Proxy & \cmark & \cmark & \no & \no & \no & \no & \text{N/A} & \yes \\
& Polymorphism & -- & \cmark & \rhalf & \yes & \yes & \rhalf & \text{N/A} & \yes \\
& Dynamic Class Loading & \cmark & \cmark & \no & \rhalf & \no & \rhalf & \text{N/A} & \yes \\
\midrule

\multirow{8}{*}{\rotatebox{90}{\textbf{C}}} 
& Function Pointers & -- & \cmark & \no & \yes & \rhalf & \rhalf & \yes & \yes \\  
& Conditional Compilation & \cmark & \cmark & \no & \rhalf & \no & \rhalf & \rhalf & \yes \\  
& External Functions & \cmark & \cmark & \no & \rhalf & \rhalf & \rhalf & \rhalf & \yes \\  
& Inline Assembly & \cmark & \cmark & \no & \no & \no & \no & \no & \yes \\  
& Macro Function Calls & \cmark & \cmark & \no & \rhalf & \no & \rhalf & \rhalf & \yes \\  
& Goto Statements & -- & \cmark & \rhalf & \yes & \yes & \yes & \yes & \yes \\  
& Setjmp/Longjmp & -- & \cmark & \no & \no & \no & \no & \no & \yes \\  
& Recursive Functions & -- & -- & \yes & \yes & \yes & \yes & \yes & \yes \\  
\bottomrule
\end{tabular}%
}
\label{tab:motivation-study}
%\vspace{-12pt}
\end{table}

\begin{figure*}[tp]
\begin{center}
\includegraphics[width=0.8\textwidth]{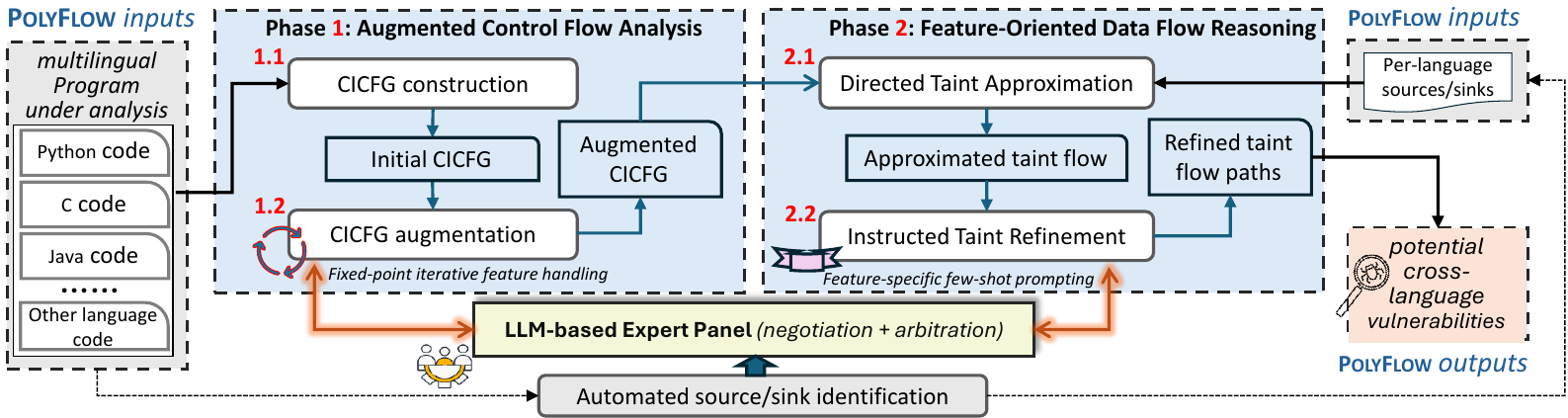}
\end{center}
%\vspace{-8pt}
\caption[Methodology Overview]{Overview of {\tech}, including its inputs, two main phases backed by an expert panel, and outputs.}
%\vspace{-10pt}
\label{fig:overview}
\end{figure*}

As summarized in Table~\ref{tab:motivation-study}, our study identified \textbf{8 C}, \textbf{6 Python}, and \textbf{5 Java} features that are complex and challenging to (hence potentially defeating) %can defeat 
static analysis. 
For each feature, we record its impact on the analysis in terms of whether, without explicitly handling it, code entities (\textbf{Ent.}) and flow facts (\textbf{Flow}) become \textit{missing} in the analysis backbone---i.e., absent nodes and edges in the underlying control-flow representation. 
For example, Python reflection/dynamic imports, Java reflection/dynamic class loading/dynamic proxies, and C inline assembly can induce both missing nodes and missing edges, as these runtime/meta-programming behaviors are not fully supported by the analyzers (with Clang being inapplicable to Java). 
As another example, Python first-class functions (and lambdas), Java polymorphic dispatch, and C function pointers primarily lead to missing edges by obscuring call targets and reachability, causing downstream flow facts to be dropped. 
From a combinatorial perspective, even if a tool is capable of cross-language analysis, it will miss holistic cross-language flow even more often---i.e., whenever any unsupported feature appears in \emph{either} language along a boundary (Python$\leftrightarrow$C or Java$\leftrightarrow$C). 
Since control flow (reachability) is a prerequisite for data flow between two program entities, missing control-flow nodes/edges implies missing data flow (hence information flow) between them.

Thus, to enable %systematic 
comprehensive 
static cross-language information flow analysis, we aim to support \textbf{all 19 features} in {\tech}. %, which combines traditional static analysis and LLMs. 
For its traditional static analysis part, 
we set three requirements to maximize applicability: 
(1) it provides direct access to the analysis backbone representations (e.g., control/data-flow graphs), 
(2) it tolerates uncompilable or incomplete code, 
and (3) it is free to use at any scale. 
(1) is also crucial for flexible customization towards greater capabilities. 
In practice, Clang is language-specific, and tools like CodeQL and Infer are build-driven, while SonarQube exposes only limited internal program representations to custom analyses. 
Only Joern meets our requirements while offering a uniform, queryable graph backbone for multi-language code. 
Accordingly, we build {\tech} on Joern, which means we still need to explicitly handle the remaining \textbf{15 out of 19} features that are not %fully 
currently supported.

%"For analysis development (vs. out-of-box scanning), Joern's CPG provides the most flexible foundation among open-source tools." - Joern: A Platform for Analyzing Binary and Source Code (ACSAC 2020)

\section{Technique}\label{sec:technique}
%This section describes the design of {\tech}. 
%our technical approach. % via the design of {\tech}. %, for analyzing cross-language programs and detecting potential vulnerabilities. 
We begin with an overview of {\tech} 
%of the framework (§\ref{overview}) 
and then detail its two key phases. %: augmented control-flow analysis (§\ref{phase 1}) and feature-oriented data-flow reasoning (§\ref{phase 2}). 
These phases work in tandem to systematically uncover cross-language information flows in \textit{the entire repository} of a given multi-language project, including those induced by the challenging language features. 
%that might otherwise evade traditional static analyses, particularly in complex multi-language settings.

\subsection{Approach Overview}\label{overview}
%{\tech} is designed to systematically identify potential vulnerabilities in Python–C programs by combining static and language-feature-oriented analyses. 
As shown in Figure~\ref{fig:overview}, {\tech} accepts two main \textbf{inputs}: the target multilingual program (project repository) %source code 
and user-specified lists of information sources and sinks. 
%These inputs feed into two major phases that work in tandem to uncover data-flow paths potentially overlooked by traditional static approaches.
Optionally, an LLM-based assisting component can be used to automatically extract sources/sinks from the project to replace or expand the lists. 
With these inputs, {\tech} works like a conventional static data flow analyzer at high level, 
starting with control flow analysis in \textbf{Phase 1}, followed by data flow analysis in \textbf{Phase 2}, 
both assisted by multiple LLMs collaborating via negotiation and arbitration (i.e., LLM-based {\em Expert Panel}). 

Specifically, \textbf{Phase 1} %, we construct the CICFG of the given system which integrates its Python and C components. Starting from a basic control-flow structure, we iteratively augment the CICFG to capture diverse features of each language, including decorators in Python and pointer manipulations in C. 
performs %intraprocedural and then interprocedural 
control flow analyses within and across the two languages, hence constructing the basic \textit{cross-language interprocedural control flow graph} (\ul{CICFG}) 
in Step \textbf{1.1}.   
It then leverages the expert panel to augment the CICFG via a fixed-point iteration of handling the targeted language features in Step \textbf{1.2}. 
This augmentation enriches the graph with additional edges and nodes, resulting in the \textit{augmented CICFG}. 
%ensuring that subtle control-flow behaviors---often missed in conventional analysis—--are as completely represented as possible. 
The goal of this phase is to lay the basis for data flow tracking in the next phase, which typically works on a control flow representation. 
The rationale is that via the augmentation CICFG may capture subtle control-flow behaviors---often missed in conventional analysis—--as completely 
as possible by leveraging the capabilities of LLMs. 

Based on this augmented CICFG, \textbf{Phase 2} aims to first narrow down the data flow analysis scope via directed taint approximation---approximating 
the taint flow with mere control flow starting from any source and exploring the CICFG as directed towards any sink (Step \textbf{2.1}). 
The rationale is that the resulting control-flow paths between the sources/sinks form a focused roadmap for the LLMs to reason about data flow as a much simplified task. 
Then, the approximation is refined through an instructed taint refinement step (\textbf{2.2}), by carefully instructing the expert panel to propagate data flow along 
the source-sink reachable control-flow paths one function at a time. 
During the propagation, {\tech} maintains accumulated tainted data and coordinates the interprocedural and inter-language context, while dealing with 
challenging language features (that induce implicit data flows) via feature-specific few-shot prompting the LLMs. 
The rationale is to minimize the LLMs' task complexities hence reducing their hallucination and maximizing their accuracy. 

%The \textbf{Phase 2} revolves around feature-oriented data-flow reasoning. Initially, we perform a directed taint approximation by extracting reachability-based subgraphs for each source–sink pair, forming a preliminary mapping of how external inputs can propagate through the program. We then refine these results through an instructed taint refinement step, which leverages language-specific insights and external function semantics to yield a more precise view of how sensitive data may traverse function and language boundaries. 
As a result, {\tech} produces a consolidated list of cross-language taint flow paths that may present security risks to be confirmed, as the framework \textbf{outputs}. 
%offering a bad for further validation or mitigation measures.

%\vspace{-6pt}
\subsection{Automated Source/Sink Identification}
Typically, a static information flow analysis takes \textit{user-specified} sources/sinks hence using each source-sink pair as a query. 
While there may be sources/sinks (e.g., common/standard language SDK APIs) to identify manually, 
it can be a daunting task for users to systematically do so for a given/arbitrary project, especially multi-language ones. 

Thus, to facilitate its practical use, {\tech} comes with an assisting component to automatically extract sources/sinks from the given multilingual project repository. 
This design is motivated by the observation that source/sink definitions are often language-specific and project-dependent, making manual specification both error-prone and non-scalable in cross-language settings.
Leveraging the expert panel, it instructs the LLMs to achieve the task using two carefully designed prompt templates (as found in our artifact package). The source identification prompt guides the LLM to recognize functions or variables that introduce external or sensitive data into the system, such as user inputs, file reads, or network requests. The sink identification prompt directs the LLM to pinpoint function calls or other operations that can potentially leak data, such as logging, file/network writes. These prompts are structured to provide the LLM with sufficient context about the codebase, enabling it to make informed decisions by leveraging the LLM's understanding of code semantics and common API patterns. 
Unlike recent tools designed for a single host language such as Java~\cite{li2025iris},
our prompts avoid hard-coding vulnerability types,
allowing the same mechanism to generalize across different host--native language pairs.

%Before constructing the CICFG and performing feature-oriented taint analysis, {\tech} carries out a pre-analysis step to identify and label potential external input points (sources) and critical output functions (sinks). This preliminary phase has two primary objectives. First, it automatically explores the program codebase to discover candidate sources and sinks through LLM queries. Second, the user may also provide a custom list of source/sink definitions, either to account for domain-specific APIs or to refine and override automatic expert panel (LLM-driven analysis, detailed in §~\ref{expert panel}) detections in cases where proprietary or unusual functions exist. {\tech} merges these user inputs with the set of sources and sinks obtained from automatic source-sink identification, ensuring that no critical entry or exit point is overlooked. With a unified source-sink register thus established, we transition to building the augmented CICFG (see §~\ref{phase 1.2}), where these marked entry and exit nodes guide the static and feature-based analyses that follow.

\subsection{LLM-based Expert Panel (Common Module)}\label{expert panel}
To enhance the robustness and cost-effectiveness of our analyses, we propose an expert panel method combining the strengths of non-reasoning (generative) and reasoning LLMs. Non-reasoning LLMs are computationally efficient but prone to instability and hallucination. Conversely, reasoning LLMs provide greater stability and accuracy but at higher costs.

Our expert panel employs cross-validation using two non-reasoning models for initial analysis. When disagreements arise, a reasoning model is then invoked as an arbiter. This combined approach capitalizes on the efficiency of non-reasoning models for most tasks, reserving the resource-intensive reasoning model for ambiguous or complex cases, thereby achieving an optimal balance between accuracy, stability, and computational expense, which allows {\tech} to scale to large repositories with analysis stability.
% as the more expensive reasoning model is invoked only when semantic ambiguity
% cannot be resolved through cross-validation.

%\vspace{-8pt}
\subsection{Augmented Control-Flow Analysis (Phase 1)}\label{phase 1}
%A program analysis is often based on a program representation, such as the control flow graph (CFG). By default, CFG is intraprocedural, representing control flow within a function. For interprocedural analysis, per-function CFGs are linked together through calling relationships between the functions, forming an interprocedural CFG (or ICFG). To enable holistic static analysis of multi-language software, a multilingual code representation is also needed, for which further link per-language ICFGs through control flow induced by cross-language functions. We refer to this representation the \textbf{cross-language interprocedural control flow graph (CICFG)}. Thus, a CICFG node represents a statement in a multilingual program, while a CICFG edge represents control flow between two nodes within and across the languages.

{\tech} constructs and refines a CICFG that spans both languages. %Python/Java and C code. 
This graph provides the foundational structure for identifying taint flows in multi-language programs. 
%We first describe how the basic CICFG is built (§~\ref{phase 1.1}), and then show how we augment this initial graph with language-specific features (§~\ref{phase 1.2}), ensuring that subtle cross-language behaviors are not lost during static analysis.

\subsubsection{CICFG Construction}\label{phase 1.1}
The lack of a ready-made, whole-program static information-flow analysis tool
for real-world cross-language systems
motivates us to adapt existing infrastructure to our needs. We start from producing a cross-language, interprocedural control-flow graph for each analyzed project. Following established practices in ICFG construction~\cite{sinha01apr}, every node in the resulting CICFG corresponds to a statement or operation in either language's portions of the codebase, and directed edges indicate the potential transfer of control. 
%Thus, a CICFG node represents a statement in a multilingual program, while a CICFG edge represents control flow between two nodes within and across the languages.
By explicitly modeling inter-language call edges via FFIs,
the CICFG establishes a unified control-flow substrate on which
subsequent language-agnostic data-flow reasoning can be performed.

The construction proceeds by first extracting intraprocedural CFGs for each function or method in the target source code files. Next, these per-method graphs are connected into an intra-language call graph, one for each language unit. We then identify cross-language call edges by examining how the two languages (e.g., Python and C) components interact, guided by FFI specifications (e.g., Python C APIs). For instance, if a C extension module provides a function accessible from Python, that interface call is represented in the graph as an inter-language edge. Finally, we merge the two intra-language call graphs, along with these cross-language edges, to form the CICFG. Each node is annotated with contextual information, such as associated function names and the source-code location (line no.), facilitating later traversal of the graph.

\subsubsection{CICFG Augmentation}\label{phase 1.2}

While the initial CICFG captures control-flow connections at the function and statement levels, cross-language programs commonly employ various features that can introduce implicit or hidden control flows. For example, Python’s decorator, reflection, and import mechanisms, along with C’s macros, function pointers, and conditional compilation, often cause static analyses to miss relevant control paths if these features are not explicitly accounted for. To address these complexities, {\tech} augments the CICFG in an iterative, feature-driven manner.
This design is motivated by the observation that language features often interact
and introduce control-flow effects transitively,
making one-shot augmentation insufficient in cross-language settings.

Initially, we scan the project for a range of language features using specialized handler functions; for example, we detect macros and function pointers in C source files, and decorators or dynamic imports in Python. When a handler discovers any such feature, it marks that file for further analysis and adds it to a global worklist. Features that are closely related---for example, macros and conditional compilation---trigger updates to each other’s state so that modifications to one construct can be properly reflected in another. The algorithm then proceeds in a fixed-point manner: if re-analysis of certain files detects new features or modifies existing ones, the corresponding handlers are re-invoked until no further updates are generated. %, as detailed in Algorithm~\ref{algo:feature_controller} in Appendix~\ref{sec:phase1algo}.

% \section{Details on Phase 1's Controller Algorithm}\label{sec:phase1algo}
% Here we present the details on the fixed-point iterative feature handling control logic, which invokes individual feature handlers (as exemplified in Appendix~\ref{sec:featurehandlingalgos-c} for function pointers in C and Appendix~\ref{sec:featurehandlingalgos-python} for decorators in Python) 
% until reaching a fixed point. 

% The handler has mainly a fixed-point iterative control logic, which invokes individual feature handlers  
% %(as exemplified in Appendix~\ref{sec:featurehandlingalgos-c} for function pointers in C and Appendix~\ref{sec:featurehandlingalgos-python} for decorators in Python) 
% until reaching a fixed point. 
The key rationale for this iterative nature of the algorithm lies in the dependencies between different features regarding their impact on the static analysis. 
For instance, for the C language, if the feature handler for function pointers discovered and resolved a reflective functional callsite, 
the same handler needs to run further to discover additional such method-level control flow through that reflective callee. 
For another example, after the conditional compilation handler resolves a conditionally compiled code block, all feature handlers for the C language should be triggered to discover and resolve any features in that code block. The iteration stops when no more change in control flow can be harvested---i.e., the CICFG augmentation process converges. 

\setlength{\textfloatsep}{0pt} \label{line:1}
\begin{algorithm}[!htbp]
\scriptsize
\caption{\footnotesize{Fixed-point iterative feature handling}}
\label{algo:feature_controller}
\SetKwProg{Fn}{Function}{}{end}
\SetKwFunction{WalkDirectory}{WalkDirectory}
\SetKw{Continue}{continue}
\SetKw{Return}{return}
\LinesNumbered

\KwIn{A project directory path \texttt{proj\_path}}
\KwOut{Analysis results of various features}

\BlankLine
all\_features $\gets$ c\_features + py\_features \; \label{l:allfeatures}
feature\_related\_mapping $\gets$ get\_featuremapping() \; \label{l:featuremapping}

startfile\_dict $\gets$ empty dictionary \; \label{l:startfile_dict}
worklist $\gets$ empty set \; \label{l:worklist}

\ForEach{\texttt{feat} in \texttt{all\_features}}{ \label{l:forFeature}
    startfile $\gets$ [] \; \label{l:startfile_init}
    analyzeFunc $\gets$ feature\_handler\_mapping[\texttt{feat}] \; \label{l:analyzeFunc}
    \ForEach{\texttt{file\_path} in \WalkDirectory(\texttt{proj\_path})}{ \label{l:walkDir}
        (startf, \_) $\gets$ analyzeFunc(\texttt{file\_path}) \; \label{l:analyzeCall}
        startfile.extend(startf) \; \label{l:startfileExtend}
    }
    startfile\_dict[\texttt{feat}] = startfile \; \label{l:startfileDictAssign}
    \If{startfile $\neq$ []}{ \label{l:ifStartfile}
        worklist = worklist $\cup$ feature\_related\_mapping[\texttt{feat}] \; \label{l:worklistUnion}
    }
}

\While{worklist \texttt{ not empty}}{ \label{l:whileWorklist}
    feat $\gets$ worklist.pop() \; \label{l:popFeat}
    startfile $\gets$ startfile\_dict(\texttt{feat}) \; \label{l:getStartfile}
    analyzeFunc $\gets$ feature\_handler\_mapping[\texttt{feat}] \; \label{l:getAnalyzeFunc}
    is\_updated $\gets$ False \; \label{l:isUpdatedInit}
    \ForEach{\texttt{file\_path} in startfile}{ \label{l:forFilePath}
        (\_, updated) $\gets$ analyzeFunc(\texttt{file\_path}) \; \label{l:analyzeAgain}
        \If{updated == True}{ \label{l:checkUpdated}
            is\_updated $\gets$ True \; \label{l:setIsUpdated}
        }
    }
    \If{is\_updated}{ \label{l:ifIsUpdated}
        worklist = worklist $\cup$ feature\_related\_mapping[\texttt{feat}] \; \label{l:worklistUnion2}
    }
}
\end{algorithm}

Specifically, in Algorithm~\ref{algo:feature_controller}, lines~\ref{l:allfeatures}-\ref{l:featuremapping} define the union of \texttt{c\_features} and \texttt{py\_features} as \texttt{all\_features}, and retrieve a \texttt{feature\_related\_mapping} that specifies dependencies among features. Lines~\ref{l:startfile_dict} and~\ref{l:worklist} initialize a dictionary for storing file lists (\texttt{startfile\_dict}) and an empty \texttt{worklist} for iterative processing. 
From lines~\ref{l:forFeature}-\ref{l:worklistUnion}, the algorithm iterates over each feature, invokes the corresponding handler on every file found in the project directory, and collects returned \texttt{startfile} information. If \texttt{startfile} is non-empty, the algorithm merges related features into the \texttt{worklist} based on \texttt{feature\_related\_mapping}. 
Lines~\ref{l:whileWorklist}-\ref{l:worklistUnion2} process features in the \texttt{worklist}. Each popped feature retrieves its file list (\texttt{startfile}) and runs the handler again. If any file is updated, the algorithm merges the dependent features into the \texttt{worklist}, ensuring that changes in one feature analysis can trigger re-analysis of related features until all dependencies are stable.

By iterating through potentially interactive or mutually dependent language constructs, we ensure that the final CICFG does not omit control-flow edges that arise from dynamic or conditional behaviors. Nodes that might otherwise appear disconnected become properly linked, and call edges that are only valid under specific macros or function pointer assignments are accurately reflected. The end result is an augmented CICFG where language features that would typically frustrate purely static approaches are consistently incorporated, providing a robust foundation for our subsequent data-flow analysis.

% \subsubsection{Feature Handling} \label{sec:feature-handling}
%
% % CICFG augmentation relies on systematically handling language-specific features, which may introduce subtle or dynamic control-flow paths. In this subsection, we illustrate this feature handling approach by detailing our method for analyzing function pointers in C, an inherently dynamic construct often challenging for static analyses.
% Before initiating taint propagation, {\tech} performs a critical pre-analysis step to account for implicit data flows caused by advanced language features. 
% Constructs such as reflection in Python or function pointer in C can create data-flow paths invisible to standard syntactic analysis. 
% To capture these, {\tech} leverages its expert panel to identify the presence of such features within a function and to infer how they implicitly propagate data between variables. 
% The resulting information is used to augment the initial CICFG by adding edges and nodes representing the previously implicit flows. 
% The detailed implementation and logic for each feature handler are provided in our supplementary artifact.

\textbf{Individual Feature Handling}.
For each of our targeted features, {\tech} has a dedicated feature handler. 
In this handler, {\tech} traverses all the feature presence units (FPUs) one at a time, which defines the minimal scope where the feature could be resolved. 
For most of the features targeted, an FPU is a function; yet for some features, it is a source file. 
When processing each FPU, {\tech} leverages the expert panel to first (1) detect the presence of the feature, followed by 
(2) identifying the missing nodes and/or edges (Table~\ref{tab:motivation-study}), and (3) when identified, it adds the nodes/edges to the CICFG accordingly. 
While for different features, the handling logic varies, the handler (including for additional features) always follows these three main steps.

%\section{Feature Handler Summary}\label{sec:feature_handler_summary}

Table~\ref{tab:feature_handler_summary} summarizes the detection cue and CICFG augmentation pattern of the 18 feature handlers in {\tech} (the 19th feature, recursive functions, is supported natively by Joern, on which the base CICFG is built). Each row instantiates the common feature handling logic %(Algorithm~\ref{algo:abstract_handler} in the main body) 
for one specific language feature; per-feature pseudocode is provided in the subsequent sections. The ``Prop?'' column marks handlers that run an inner worklist to chase chain effects to convergence.

\renewcommand{\arraystretch}{1.1}
\begin{table*}[!htbp]
\centering
\scriptsize
\caption{Per-feature instantiation of the common feature handling logic.}
\label{tab:feature_handler_summary}
\resizebox{\textwidth}{!}{%
\begin{tabular}{|l|l|l|c|}
\hline
\textbf{Feature (Lang.)} & \textbf{Detection cue} & \textbf{CICFG augmentation pattern} & \textbf{Prop?} \\
\hline
Function Pointers (C)         & function-pointer typed expr               & callsite $\to$ bound target entry $\to$ callsite           & Y \\
\hline
Conditional Compilation (C)   & preprocessor \texttt{\#ifdef}/\texttt{\#endif} directive & per-branch nodes + intra-/inter-branch edges & N \\
\hline
Macros to Function Calls (C)  & macro expanding into invocation           & expanded callsite $\to$ callee entry $\to$ callsite        & N \\
\hline
Inline Assembly (C)           & \texttt{asm}/\texttt{\_\_asm} block       & asm-ref node + side-effect edges                            & N \\
\hline
Goto Statements (C)           & \texttt{goto} keyword and label           & label-jump intra-procedural edges                           & N \\
\hline
Setjmp/Longjmp (C)            & \texttt{setjmp}/\texttt{longjmp} calls    & longjmp-site $\to$ matching setjmp $\to$ continuation       & N \\
\hline
Dynamic Linking (C)           & \texttt{dlopen}/\texttt{dlsym}            & resolved-symbol $\to$ target entry $\to$ resolve-site       & N \\
\hline
Decorators (Python)           & \texttt{@decorator} syntax above def       & callsite $\to$ wrapper entry $\to$ original-call resumption & N \\
\hline
First-Class Functions (Python)& function passed as value or stored        & callsite $\to$ resolved target entry $\to$ callsite         & Y \\
\hline
Lambda Expressions (Python)   & \texttt{lambda} keyword                   & exec-site $\to$ lambda body $\to$ exec-site                  & N \\
\hline
Dynamic Typing (Python)       & duck-typed call / \texttt{getattr}        & resolved-method entry $\to$ call-return                      & N \\
\hline
Reflection (Python)           & \texttt{getattr}/\texttt{importlib}/\texttt{\_\_class\_\_} & resolved callsite $\to$ reflect target $\to$ callsite & N \\
\hline
Dynamic Imports (Python)      & \texttt{\_\_import\_\_}/\texttt{importlib.import\_module} & import-site $\to$ module entry & N \\
\hline
Reflection (Java)             & \texttt{Method.invoke} / \texttt{Class.forName} & callsite $\to$ resolved member $\to$ callsite           & N \\
\hline
Lambda Expressions (Java)     & \texttt{->} lambda or method reference    & exec-site $\to$ lambda body $\to$ exec-site                  & N \\
\hline
Dynamic Proxy (Java)          & \texttt{Proxy.newProxyInstance}           & proxy callsite $\to$ \texttt{InvocationHandler.invoke} $\to$ callsite & N \\
\hline
Dynamic Class Loading (Java)  & \texttt{Class.forName} / \texttt{ClassLoader.load} & loadsite $\to$ \texttt{<clinit>} $\to$ loadsite        & Y \\
\hline
Polymorphism (Java)           & virtual call site / \texttt{instanceof}   & virtual call $\to$ concrete impl entry $\to$ callsite        & Y \\
\hline
\end{tabular}
}
\end{table*}

%For space limit, 
We have provided the detailed algorithm and explanations for each feature handler in our artifact package. 
Appendix~\ref{sec:featurehandlingalgos-c},~\ref{sec:featurehandlingalgos-python}, and~\ref{sec:featurehandlingalgos-java} provide examples of such feature handling algorithms for Python, Java, and C, respectively. 

\subsection{Feature-Oriented Data Flow Reasoning (Phase 2)}\label{phase 2}
Having constructed and augmented the CICFG, we next detect and refine taint flows that may traverse multiple language boundaries in the presence of complex language features. {\tech} orchestrates this phase in two steps: directed taint approximation stage (§~\ref{phase 2.1}) that enumerates candidate flows from sources to sinks, and instructed refinement stage (§~\ref{phase 2.2}) that uses feature knowledge and LLM-assisted context to prune or expand these flows as necessary. The final result is a set of potential vulnerabilities, each represented by a path from source to sink across the augmented control-flow graph.

%\vspace{-6pt}
\subsubsection{Directed Taint Approximation}\label{phase 2.1}
The first step of our data-flow reasoning phase applies a control-flow-reachability-based analysis on top of the augmented CICFG. We either use the user-specified source/sink lists or begin by collecting pairs of taint sources and sinks. For each such pair 
⟨\textit{src},\textit{sink}⟩, we slice the augmented CICFG to extract the subgraph of control-flow nodes lying on the possible paths from \textit{src} to \textit{sink}. This subgraph reflects an initial mapping of how data originating from \textit{src} could propagate through both Python and C components until reaching \textit{sink}.

We then aggregate these paths at purely-control-flow level, a process that highlights the subsets of code relevant to each source-sink traversal. By operating on a control-flow-based approximation,
we deliberately trade early precision for scalability.
This design ensures that expensive, feature-aware reasoning
is only applied to source--sink pairs that are structurally feasible,
which is crucial for analyzing large cross-language repositories practically. Once this reachability-based approximation is complete, it yields a set of candidate paths for each ⟨\textit{src},\textit{sink}⟩ pair, awaiting further scrutiny in the next step.

%\vspace{-6pt}
\subsubsection{Instructed Taint Refinement}\label{phase 2.2}
After gathering all candidate paths, we refine them through a deeper inspection that leverages both static feature handlers and LLM-assisted analyses. For every subgraph derived from the source–sink pairs, we traverse its constituent functions one by one, carrying along a set of tainted variables (initially those introduced at the source).
At each function, we invoke specialized handlers designed to account for language-specific constructs, such as value transfer via pointer indirection in C. The rationale is to align the granularity of reasoning with
the locality of language features,
allowing LLM assistance to be tightly scoped and context-aware. 
These handlers systematically update the set of tainted variables, capturing implicit or indirect data flows that might be missed by purely syntactic analysis. Once the feature handlers finish, %have run, %we engage 
the expert panel confirms or %further 
extends the taint set. The prompt sent to the expert panel encapsulates the current function’s code snippet and any already tainted variables, asking for potential aliases, hidden references, or dynamic behaviors. The expert panel’s response may identify additional variables as tainted, which are merged back into our working set of tainted data.

In scenarios where a function passes tainted variables across language boundaries—for instance, when Python invokes a C extension function or when C calls back into Python—{\tech} consults prior knowledge of cross-language APIs and the rules established in CICFG. If these variables reach a sink node, we record a fully realized taint flow spanning the entire path from source to sink. Once all subgraphs have been traversed, the system produces a consolidated view of verified vulnerabilities, each annotated with the final chain of tainted variables and functions that lead to an external output. By merging feature-driven insights with LLM guidance, we gain a precise and robust perspective on taint propagation—even in multi-language programs with advanced or unusual language constructs. %The full details are provided in Appendix~\ref{sec:phase2algo}.

 %\section{Details on Phase 2 Algorithm}\label{sec:phase2algo}

\begin{algorithm}[!htbp]
\scriptsize
\caption{\footnotesize{Feature-oriented data flow reasoning}}
\label{algo:llmTaintAnalysisFeatures}
\SetKwProg{Fn}{Function}{}{end}
\SetKwFunction{BuildSlice}{BuildSlice}
\SetKwFunction{GetSubgraphFunctions}{GetSubgraphFunctions}
\SetKwFunction{SendToLLM}{SendToLLM}
\SetKwFunction{GetFeatureHandlers}{GetFeatureHandlers}
\SetKwFunction{AnalyzeFunctionWithFeature}{AnalyzeFunctionWithFeature}
\SetKwFunction{GetNextFunctionAndArgs}{GetNextFunctionAndArgs}
\SetKw{Break}{break}
\SetKw{Return}{return}
\LinesNumbered

\KwIn{A set of source/sink pairs $S$, a Control-Integrated Control Flow Graph $CICFG$}
\KwOut{Taint analysis results with updated tainted variables per function}

% Instead of \Fn{\LLMAnalysisWithFeatures{$S,\,CICFG$}}, we do:
\Fn{LLMAnalysisWithFeatures($S,\, CICFG$)}{
    \ForEach{$(src,\, sink) \in S$}{ \label{algoLine1}
        subgraph $\gets$ \BuildSlice{$CICFG,\; src,\; sink$}\; \label{algoLine2}
        functionsInSubgraph $\gets$ \GetSubgraphFunctions{$subgraph$}\; \label{algoLine3}

        taintedVars $\gets$ set of variables initially tainted by $src$\; \label{algoLine4}
        features $\gets$ \GetFeatureHandlers{}\; \label{algoLine5}
        \tcp{features represents a set of feature handlers for implicit data flow, etc.}

        \ForEach{$currentFunc \in functionsInSubgraph$}{ \label{algoLine6}
            \tcp{(1) Update taintedVars using feature handlers}
            \ForEach{$feature \in features$}{ \label{algoLine7}
                newTaintedVarsFromFeature $\gets$ \AnalyzeFunctionWithFeature{feature,\; currentFunc,\; taintedVars}\; \label{algoLine8}
                taintedVars $\gets$ taintedVars $\cup$ newTaintedVarsFromFeature\; \label{algoLine9}
            }

            \tcp{(2) Ask the LLM for additional taint information}
            codeSnippet $\gets$ code of $currentFunc$ from $subgraph$\; \label{algoLine10}
            promptCode $\gets$ "Given this function code and current tainted vars, list all tainted vars in this function:"
                + codeSnippet + \text{" Tainted vars:"} + taintedVars\; \label{algoLine11}
            llmResTaint $\gets$ \SendToLLM{$\text{promptCode}$}\; \label{algoLine12}
            newTaintedFromLLM $\gets$ parse($\text{llmResTaint}$) into a set of tainted variables\; \label{algoLine13}
            taintedVars $\gets$ taintedVars $\cup$ newTaintedFromLLM\; \label{algoLine14}

            \tcp{(3) Determine and pass tainted vars to the next function}
            (nextFunc, passedTaintedVars) $\gets$ \GetNextFunctionAndArgs{currentFunc,\; subgraph,\; taintedVars}\; \label{algoLine15}
            \If{$\text{nextFunc} = \text{NULL}$}{ \label{algoLine16}
                \tcp{No further propagation, end loop for this subgraph}
                \Break
            }
            \Else{ \label{algoLine17}
                taintedVars $\gets$ passedTaintedVars\; \label{algoLine18}
            }
        }
    }
    \Return taintedVars\; \label{algoLine19}
}
\end{algorithm}

As shown in Algorithm~\ref{algo:llmTaintAnalysisFeatures}, {\tech}, for each $(src, sink)$ pair in the input set $S$, first builds a subgraph (denoted \emph{subgraph}) that captures paths from \emph{src} to \emph{sink} in the CICFG, and then extracts all functions appearing in that subgraph (Lines~\ref{algoLine1}--\ref{algoLine3}).
Initialize a set of tainted variables (\emph{taintedVars}) based on the source $src$, and retrieve a collection of feature handlers (\emph{features}). These handlers can account for additional data-flow scenarios such as implicit flows or specific language constructs (Lines~\ref{algoLine4}--\ref{algoLine5}).
For each function in the subgraph, iterate over every feature handler to identify and add new tainted variables (\emph{newTaintedVarsFromFeature}) to the existing \emph{taintedVars} (Lines~\ref{algoLine6}--\ref{algoLine9}).
Use the code of the current function (\emph{codeSnippet}) to query the expert panel. The expert panel response may introduce further tainted variables (\emph{newTaintedFromLLM}). Merge these into \emph{taintedVars} (Lines~\ref{algoLine10}--\ref{algoLine14}).
Determine which tainted variables, if any, should be passed to the next function. If there is no next function (\emph{nextFunc} = NULL), end the loop for this subgraph. Otherwise, update \emph{taintedVars} accordingly and continue (Lines~\ref{algoLine15}--\ref{algoLine18}).
After analyzing all \emph{(src, sink)} pairs, return the resulting tainted variables. This final tainted variable set can then be used for further security analysis or vulnerability detection (Line~\ref{algoLine19}).

%\subsubsection{Feature Handling}\label{phase 2.3}
\textbf{Individual Feature Handling}.
Before initiating taint propagation, we explicitly handle language-specific features that might introduce implicit or indirect data flows. Certain advanced constructs inherent to Python and C—such as dynamic reflection, protocol implementations, and magic methods in Python, or unions, type punning, and volatile or atomic operations in C—can obfuscate or complicate data-flow analysis. Detecting these implicit flows early ensures improved accuracy and completeness in the subsequent taint analysis in our framework.

To capture such implicit data flows, {\tech} employs an expert-driven approach leveraging LLM assistance. For each function within the augmented CICFG, we first invoke the expert panel to detect the presence of these complex language features. Given a function's code snippet, the panel identifies whether it utilizes any implicit-flow-causing features. 
If any such features are detected, the expert panel further determines how these constructs implicitly propagate tainted data. Specifically, the panel infers which additional variables might become tainted through the identified implicit flows, supplementing our initial tainted variable set. The results from this expert-assisted analysis are integrated into the initial taint sets for each function.
By proactively addressing implicit flows at the outset of our analysis, we significantly enhance the precision and effectiveness of subsequent taint propagation steps. This structured handling of complex language features equips {\tech} with a robust foundation for detecting vulnerabilities.

%\vspace{-2pt}
\section{Implementation and Limitations}\label{sec:implementation}

This section details the implementation of {\tech} %, emphasizing its extensible architecture, and 
and 
discusses its current %technical and implementation-specific 
limitations.

%\vspace{-2pt}
\subsection{Implementation}
{\tech} is implemented with a modular architecture for future enhancements and adaptability, whch allows for:
% , though it currently possesses certain technical and implementation-specific limitations. The framework's design distinctly separates concerns, facilitating extensibility. 
% Core analysis phases like CICFG augmentation and feature-oriented data-flow reasoning are orchestrated to work with abstract representations and defined interfaces. 
% This allows for:

\textbf{Broader Language Support.} New language pairs can be incorporated by developing parsers to generate the initial control-flow structures and new language-specific “feature handlers". The underlying analysis logic remains largely independent of specific language frontends.

\textbf{Interchangeable LLM Components.} The “Expert Panel"  and other LLM interactions are accessed through an interface that can be adapted to different foundation models, allowing for upgrades or changes in LLM providers or types.

\textbf{Alternative Static Analysis Infrastructure.} While Joern is currently used for initial intraprocedural CFG generation, {\tech}’s subsequent CICFG construction and augmentation logic are self-contained. This permits the potential integration of alternative static analysis tools for the initial graph generation phase if with compatible input format.

\begin{table}[tp]
\caption{Summary of xFlowBench}
\centering
% \footnotesize
\resizebox{1\columnwidth}{!}{
\begin{tabular}{clccccc}
\hline
\multirow{2}{*}{\textbf{Lang.}} & \multirow{2}{*}{\textbf{Feature}} & \multirow{2}{*}{\textbf{\begin{tabular}[c]{@{}c@{}}\#micro-\\ benchs\end{tabular}}} & \multicolumn{2}{c}{\textbf{\#data points}} & \multirow{2}{*}{\textbf{\#FNs}} & \multirow{2}{*}{\textbf{\#FPs}} \\ \cline{4-5}
 &  &  & \textbf{\#edges} & \textbf{\#nodes} &  &   \\ \hline \hline
\multirow{6}{*}{\rotatebox{90}{\textbf{Python}}} 
 & Decorators & 5 & 24 & 2 & 1 & 0 \\
 & First-Class Functions & 4 & 9 & 1 & 0 & 0 \\
 & Lambda Functions & 6 & 11 & 7 & 2 & 0 \\
 & Dynamic Typing & 4 & 8 & 1 & 0 & 0 \\
 & Reflection & 7 & 9 & 2 & 3 & 0 \\
 & Dynamic Imports & 6 & 4 & 2 & 0 & 0 \\ \hline
\multirow{5}{*}{\rotatebox{90}{\textbf{Java}}} 
 & Reflection & 7 & 14 & 3 & 1 & 0 \\
 & Lambda Expression & 7 & 17 & 4 & 1 & 0 \\
 & Dynamic Proxy & 7 & 12 & 2 & 0 & 0 \\
 & Polymorphism & 7 & 16 & 3 & 0 & 0 \\
 & Dynamic Class Loading & 6 & 10 & 2 & 1 & 0 \\ \hline
\multirow{8}{*}{\rotatebox{90}{\textbf{C}}} 
 & Function Pointers & 5 & 20 & 2 & 0 & 0 \\
 & Conditional Compilation & 4 & 4 & 2 & 0 & 0 \\
 & External Functions & 5 & 4 & 2 & 3 & 0 \\
 & Inline Assembly & 5 & 7 & 1 & 3 & 0 \\
 & Macro Function Calls & 6 & 6 & 1 & 1 & 0 \\
 & Goto Statements & 6 & 9 & 0 & 0 & 0 \\
 & Setjmp/Longjmp & 5 & 8 & 0 & 0 & 0 \\
 & Recursive Functions & 2 & 6 & 2 & 0 & 0 \\ \hline
\multicolumn{2}{c}{\textbf{Total}} & \textbf{104} & \textbf{198} & \textbf{39} & \textbf{16} & \textbf{0} \\ \hline
\end{tabular}
}
\label{tab:mb}
%\vspace{-16pt}
\end{table}

%\vspace{-4pt}
\subsection{Limitations}
%\vspace{-2pt}
%Despite its capabilities, {\tech} has the following limitations:
\textbf{Reliance on LLM Efficacy.} The accuracy of identifying challenging language features and reasoning about data flows is dependent on the capabilities of the integrated LLMs. While mitigations like the expert panel are in place, LLM-based analysis can be susceptible to inaccuracies or “hallucinations," and the interpretability of LLM decisions can be challenging.

\textbf{Dependence on Source/Sink Accuracy.} The overall analysis is sensitive to the quality of the initial source and sink definitions, whether provided by users or identified via LLM-based methods. Inaccuracies here can impact the relevance and completeness of detected taint flows.

%\textbf{Current Language Pair Focus.} The current {\tech} implementation is specifically for Python-C systems. Extending it to other language combinations requires considerable effort, including parsing new foreign function interfaces and developing feature handlers for the new languages.

\textbf{Multilingual Construction Focus.} Our implementation of {\tech} currently only supports Python-C and Java-C.
% , given its top popularity and impact in multi-language domains (mainly due to AI/ML systems built on them). 
Moreover, we currently focus on FFI as the interfacing mechanism between Python and C. Although FFI represents the dominant mechanism, information flow induced by non-FFI interfacing would not be captured by our tool.

%\vspace{-6pt}
\section{\textit{xFlowBench}: A Multilingual Microbench}\label{sec:microbench}
% \usepackage{multirow}

% \usepackage{multirow}

% Evaluating the precision and recall of a cross-language static analysis like {\tech}, especially its handling of complex or dynamic language features, requires a suitable benchmark suite. 

For evaluating the precision and recall of a cross-language static analysis like {\tech}, existing benchmarks often lack coverage for specific cross-language interactions or challenging features relevant to static analysis. Therefore, we created xFlowBench, a micro-benchmark suite tailored for evaluating static analysis across Python/Java-C boundaries.

xFlowBench comprises 104 micro-benchmarks specifically designed to test the handling of challenging language features in both Python/Java/C that can affect information flow analysis. As detailed in Table~\ref{tab:mb}, each benchmark focuses on one or more of these features, presenting scenarios where traditional static analysis might struggle. This suite allows for a systematic assessment of {\tech}'s ability to correctly identify and handle these complex constructs during its cross-language information flow analysis.

\section{Evaluation} \label{sec:eval}

We evaluated the effectiveness of {\tech} on both benchmarks and real-world Python-C and Java-C projects. Our evaluation aims to answer five research questions:

\begin{itemize}[noitemsep,topsep=2pt]
    \item[\textbf{RQ1}] How prevalent are targeted Python/Java/C features in real-world Python-C and Java-C projects?
    \item[\textbf{RQ2}]  How effective is {\tech} for static cross-language information flow analysis against SOTA baselines? 
    \item[\textbf{RQ3}]  Does each of its components contribute significantly to {\tech}'s performance? %Ablation studies.
    \item[\textbf{RQ4}]  How efficient is {\tech} on real-world systems? %What is the runtime overhead of {\tech}.
    \item[\textbf{RQ5}]  Can {\tech} find cross-language vulnerabilities? %Case studies.
    %\item[\textbf{RQ5}]  How effectively {\tech} can detect single-language language bugs in Python and C?
\end{itemize}

\textbf{Experimental Setup}.
In addition to xFlowBench, we also evaluated {\tech} against 14 real-world Python-C projects and 15 Java-C projects. Table~\ref{tab:benchmarks} shows the project size in thousands of lines of code (2nd column) and the percentage of code written in each language (3rd columns).

We conduct our experiments on a %high-performance 
workstation equipped with an AMD Ryzen Threadripper 3970X CPU, an Nvidia GeForce RTX 3090 GPU, and 256GB of RAM.
For static code analysis tasks, we utilize Joern version 4.0.216. In experiments involving the expert panel, we set the temperature parameter to 0 for LLMs.
% , ensuring deterministic behavior. 
% This configuration guarantees consistent output generation from identical inputs, thus enhancing reproducibility.
% All other environment details, libraries, and specific dependencies remain consistent across all experiments, ensuring controlled and reproducible conditions.

% We evaluate {\tech} against xx real-world multilingual systems written in Python and C as primary languages.

% \begin{table}[]
%     \centering
%     \caption{Real-world multilingual subjects}
%     \begin{tabular}{lrrr}
%         \toprule
%         Benchmark & LOC (k) & Python code \% & C code \% \\
%         \midrule
%         {\Bounter}        & 3.5   & 50.9\% & 48.2\% \\
%         {\Ultrajson}      & 5.1   & 34.3\% & 64.8\% \\ 
%         {\Simplejson}     & 6.2   & 59.8\% & 37.6\% \\
%         {\Libsmbios}      & 8.3   & 30.4\% & 64.2\% \\
%         {\Japronto}       & 9.4   & 48.2\% & 50.4\% \\
%         {\Pycurl}         & 14.6  & 54.8\% & 40.7\% \\
%         {\Msgpack}        & 15.1  & 48.7\% & 50.1\% \\
%         {\Bottleneck}     & 16.9  & 49.5\% & 48.6\% \\
%         {\Pygit}          & 17.0  & 44.6\% & 57.4\% \\
%         {\Aubio}          & 42.9  & 25.4\% & 73.3\% \\ 
%         {\Cvxopt}         & 56.0  & 39.0\% & 60.8\% \\
%         {\Pytables}       & 219.8 & 46.6\% & 52.1\% \\
%         {\Tink}           & 257.7 & 7.2\%  & 33.5\% \\ 
%         {\Pyo}            & 259.1 & 48.8\% & 50.8\% \\
%         \bottomrule
%     \end{tabular}
%     \label{tab:benchmarks}
% \end{table}

\begin{table}[t]
    \centering
    \vspace{4pt}
    \caption{Real-world multilingual subjects}
    \label{tab:benchmarks}
    \resizebox{\columnwidth}{!}{%
    \begin{tabular}{l r l r r}
        \toprule
        Benchmark & LOC (k) & Language & \#Stars & \#Forks \\
        \midrule
        {\Bounterall (Bou)}        & 3.5   & Python: 50.9\%, C: 48.2\% & 934 & 46 \\
        {\Ultrajsonall (Ult)}      & 5.1   & Python: 34.3\%, C: 64.8\% & 4,474 & 373 \\ 
        {\Simplejsonall (Sim)}     & 6.2   & Python: 59.8\%, C: 37.6\% & 1,702 & 351 \\
        {\Libsmbiosall (Lib)}      & 8.3   & Python: 30.4\%, C: 64.2\% & 214 & 40 \\
        {\Japrontoall (Jap)}       & 9.4   & Python: 48.2\%, C: 50.4\% & 8,561 & 569 \\
        {\Pycurlall (Pyc)}         & 14.6  & Python: 54.8\%, C: 40.7\% & 1,146 & 322 \\
        {\Msgpackall (Msg)}        & 15.1  & Python: 48.7\%, C: 50.1\% & 2,061 & 239 \\
        {\Bottleneckall (Bot)}     & 16.9  & Python: 49.5\%, C: 48.6\% & 1,161 & 114 \\
        {\Pygitall (Pyg)}          & 17.0  & Python: 44.6\%, C: 57.4\% & 1,712 & 402 \\
        {\Aubioall (Aub)}          & 42.9  & Python: 25.4\%, C: 73.3\% & 3,624 & 409 \\ 
        {\Cvxoptall (Cvx)}         & 56.0  & Python: 39.0\%, C: 60.8\% & 1,028 & 213 \\
        {\Pytablesall (Pyt)}       & 219.8 & Python: 46.6\%, C: 52.1\% & 1,360 & 281 \\
        {\Tinkall (Tin)}           & 257.7 & Python: 7.2\%,  C: 33.5\% & 13,565 & 1,183 \\ 
        {\Pyoall}            & 259.1 & Python: 48.8\%, C: 50.8\% & 1,407 & 145 \\
        \midrule
        {\ImBlockerFabricall(ImB)}     & 1.5   & Java: 52.1\%, C: 19.1\% & 87   & 8    \\
        {\Androidgifdrawableall (And)}  & 8.2   & Java: 38.6\%, C: 31.6\% & 9653 & 1787 \\
        {\Termuxxall (Ter)}             & 13.4  & Java: 37.2\%, C: 44.0\% & 3,425 & 480  \\
        {\Jepall (Jep)}                 & 22.0  & Java: 25.4\%, C: 51.7\% & 1,471 & 162  \\
        {\Hinnall (Hin)}                & 37.0  & Java: 10.7\%, C: 81.8\% & 640  & 133  \\
        {\Pijvall (Pij)}                & 38.6  & Java: 81.4\%, C: 4.5\%  & 343  & 71    \\
        {\Junixsocketall (Jun)}         & 60.5  & Java: 62.2\%, C: 12.0\% & 472  & 116  \\
        {\PojavLauncherall (Poj)}       & 65.5  & Java: 50.3\%, C: 24.9\% & 8,665 & 1819 \\
        {\Wolfssljniall (Wol)}          & 67.0  & Java: 66.0\%, C: 25.4\% & 68   & 40   \\
        {\Espeakngall (Esp)}            & 64.0  & Java: 6.4\%,  C: 70.1\% & 6,099 & 1,183 \\
        {\Themisall (The)}              & 135.3 & Java: 3.0\%,  C: 27.2\% & 1,947 & 156  \\
        {\Openjpegall (Ope)}            & 188.4 & Java: 1.1\%,  C: 86.2\% & 1,071 & 498  \\
        {\Turbovncall (Tur)}            & 290.2 & Java: 15.8\%, C: 78.8\% & 966  & 156  \\
        {\Mdsplusall (Mds)}             & 669.7 & Java: 36.4\%, C: 32.1\% & 90   & 51   \\
        {\Lwjglall (Lwj)}               & 1363.4& Java: 65.9\%, C: 17.9\% & 5,249 & 680  \\
        \bottomrule
    \end{tabular}%
    }
    \vspace{10pt}
\end{table}

\textbf{Baseline}.
%
% MultiQL is a prototype extension of CodeQL, a declarative static analysis framework developed by GitHub, for multilingual program analysis.  As a query-based analysis tool, MultiQL first transforms C and Python source code into databases of facts separately. These databases are subsequently merged by establishing interlanguage dataflow edges, connecting arguments in foreign function calls to their corresponding parameters in cross-language target functions. Once the merged database is generated, MultiQL performs reachability analysis, accepting user-defined sources and sinks as query inputs and computing the existence of dataflow between them as outputs. 
%
%
% \subsubsection{Dynamic Methods}
% PolyCruise~\cite{wen22usenixsecurity} is a dynamic information flow analysis (DIFA) technique for multi-language programs. It combines language-specific static analysis to compute symbolic dependencies with language-agnostic online dynamic analysis, overcoming language heterogeneity. We compare {\tech} against PolyCruise on the source/sink pairs PolyCruise used.
%
% PolyFuzz~\cite{wen23usenixsecurity} is a greybox fuzzer that holistically fuzzes multi-language systems. It uses cross-language coverage feedback and models semantic relationships between program inputs and branch predicates across languages. We compare {\tech} against PolyFuzz by comparing the ultimately found bugs.
%
%
%
We chose xLoc~\cite{yang2024learning} as a pure deep learning-based approach to the same problem. This comparison allows us to assess the benefits of our neuro-symbolic framework against an end-to-end learning solution. 

% xLoc is a deep learning-based technique for detecting and localizing multilingual bugs using only source code. It employs a customized Transformer model, pre-trained to understand multilingual control-flow structures and fine-tuned using cross-language API vicinity information to specifically target multilingual bugs. The implementation focuses on Python-C software. We compare {\tech} against xLoc on the ultimately found bugs.

\textbf{Metrics and Procedure}.
For RQ1 (feature prevalence), we measure how targeted Python/Java/C features are distributed in real-world Python-C/Java-C projects studied.

For RQ2 (effectiveness measurement), we measure the precision of sources and sinks {\tech} identified and the taint flow paths found correspondingly.

%total number of unique vulnerabilities discovered by {\tech} compared to baseline tools (MultiQL, PolyCruise, PolyFuzz, and xLoc), reporting precision where applicable.

For RQ3 (ablation study), we created several variants of {\tech} by selectively disabling key components. We compare these variants against the complete {\tech} to quantify the contribution of each component.

For RQ4 (performance evaluation), we measured runtime performance and resource consumption. Specifically, we tracked analysis time and peak memory usage.

%and scalability factors including lines of code analyzed per minute and the relationship between codebase size and analysis time.

For RQ5 (real-world bug discovery), we conducted detailed qualitative analysis of selected bugs discovered by {\tech} and case studies.

% Each case study documents the vulnerability type, how it manifests across language boundaries, why existing tools failed to detect it, and the path through which {\tech} successfully identified it.

\begin{table*}[t]
\caption{Source/Sink precision and count by project}
\label{tab:source_sink_precision_merged}
\centering
\resizebox{\textwidth}{!}{%
\begin{tabular}{ll|*{14}{c}|c|*{15}{c}|c}
\toprule
& 
& \Aubio & \Bottleneck & \Bounter & \Cvxopt & \Japronto & \Libsmbios & \Msgpack & \Pycurl & \Pygit & \Pytables & \Pyo & \Simplejson & \Tink & \Ultrajson & \textbf{Avg}
& \Lwjgl & \Mdsplus & \Pijv & \PojavLauncher & \Turbovnc & \Themis & \Espeakng & \Junixsocket & \Androidgifdrawable & \ImBlockerFabric & \Jep & \Hinn & \Termuxx & \Openjpeg & \Wolfssljni & \textbf{Avg} \\
\midrule
\multirow{2}{*}{\textbf{Source}} & \multicolumn{1}{c|}{\textbf{Precision}}
& 72\% & 74\% & 79\% & 81\% & 88\% & 78\% & 80\% & 85\% & 82\% & 78\% & 90\% & 76\% & 73\% & 86\% & \textbf{80\%}
& 78\% & 82\% & 86\% & 81\% & 82\% & 93\% & 85\% & 72\% & 82\% & 92\% & 79\% & 84\% & 81\% & 80\% & 93\% & \textbf{83\%} \\
& \multicolumn{1}{c|}{\textbf{Count}}
& 351 & 82 & 24 & 375 & 250 & 210 & 51 & 222 & 190 & 909 & 666 & 46 & 595 & 44 & \textbf{305}
& 566 & 90 & 76 & 104 & 83 & 132 & 55 & 146 & 97 & 24 & 265 & 62 & 153 & 125 & 118 & \textbf{140} \\
\midrule
\multirow{2}{*}{\textbf{Sink}} & \multicolumn{1}{c|}{\textbf{Precision}}
& 71\% & 100\% & 100\% & 100\% & 100\% & 94\% & 100\% & 94\% & 85\% & 100\% & 97\% & 100\% & 100\% & 100\% & \textbf{95\%}
& 91\% & 90\% & 88\% & 100\% & 89\% & 100\% & 83\% & 94\% & 91\% & 100\% & 97\% & 100\% & 94\% & 93\% & 92\% & \textbf{93\%} \\
& \multicolumn{1}{c|}{\textbf{Count}}
& 51 & 12 & 3 & 14 & 14 & 33 & 4 & 17 & 13 & 144 & 65 & 3 & 45 & 12 & \textbf{32}
& 64 & 10 & 8 & 12 & 9 & 15 & 6 & 16 & 11 & 5 & 30 & 7 & 17 & 14 & 13 & \textbf{16} \\
\bottomrule
\end{tabular}%
}
\end{table*}

\begin{table*}[t]
\caption{Effectiveness of {\tech} versus xLoc}
\label{tab:detection_performance}
\centering
\resizebox{\textwidth}{!}{%
\begin{tabular}{ll|*{30}{c}}
\hline
\multicolumn{2}{c|}{\textbf{}} 
& \Aubio & \Bottleneck & \Bounter & \Cvxopt & \Japronto & \Libsmbios & \Msgpack & \Pycurl & \Pygit & \Pyo & \Pytables & \Simplejson & \Tink & \Ultrajson
& \Lwjgl & \Mdsplus & \Pijv & \PojavLauncher & \Turbovnc & \Themis & \Espeakng & \Junixsocket & \Androidgifdrawable & \ImBlockerFabric & \Jep & \Hinn & \Termuxx & \Openjpeg & \Wolfssljni
& \textbf{Total} \\
\hline
\multirow{3}{*}{\textbf{{\tech}}}
& \textbf{\#Paths}
& 27 & 26 & 28 & 1 & 24 & 51 & 3 & 4 & 41 & 0 & 0 & 24 & 93 & 13
& 78 & 10 & 8 & 12 & 9 & 16 & 5 & 18 & 11 & 0 & 35 & 6 & 19 & 15 & 14
& \textbf{591} \\
& \textbf{\#TPs}
& 27 & 22 & 28 & 1 & 21 & 31 & 3 & 3 & 35 & 0 & 0 & 24 & 65 & 12
& 54 & 7 & 6 & 7 & 6 & 11 & 3 & 15 & 7 & 0 & 26 & 3 & 15 & 12 & 10
& \textbf{454} \\
& \textbf{Prec}
& 100\% & 85\% & 100\% & 100\% & 88\% & 61\% & 100\% & 75\% & 85\% & - & - & 100\% & 70\% & 92\%
& 69\% & 70\% & 75\% & 58\% & 67\% & 69\% & 60\% & 83\% & 64\% & - & 74\% & 50\% & 79\% & 80\% & 71\%
& \textbf{76.8\%} \\
\hline
\textbf{xLoc} & \textbf{}
& - & - & - & - & - & - & - & - & - & - & - & - & - & -
& - & - & - & - & - & - & - & - & - & - & - & - & - & - & -
& - \\
\hline
\end{tabular}%
}
%\vspace{-6pt}
\end{table*}

% \subsection{RQ1: How prevalent are targeted Python/C features in real-world C-Python projects?}
\subsection{RQ1: Feature Prevalence}
% To comprehensively evaluate {\tech}, we selected a diverse dataset of 14 real-world, open-source multilingual projects. 
% These projects, which primarily use a Python-C architecture, were chosen to represent a wide range of application domains and scales, from small utilities to large-scale data processing and cryptography libraries. 
% This diversity ensures that our evaluation assesses {\tech}'s performance across various coding styles, complexities, and system sizes. 
% A detailed list of these subject projects, along with their respective lines of code and the proportional distribution of Python and C code, is presented in Table~\ref{tab:benchmarks}.

To understand the landscape in which {\tech} operates, we first investigated the prevalence of several targeted challenging Python/Java/C language features within our dataset projects in Table~\ref{tab:benchmarks}. 
The presence of these features was determined by utilizing {\tech}’s intrinsic feature detection capabilities, as detailed in Section~\ref{phase 2.2}. 
% These capabilities employ specialized handlers to identify language constructs within the codebase of the subject projects, ensuring our prevalence study is grounded in the same understanding of features that {\tech} uses for its vulnerability analysis. 
%The results of this feature prevalence analysis are in the Appendix~\ref{sec:feature}.

%\section{Details on Feature Prevalence}\label{sec:feature}

\begin{table}[htbp]
\centering
\caption{Prevalence of targeted features in Python-C subjects}
\label{tab:prevalence}
\resizebox{0.5\textwidth}{!}{%
\begin{tabular}{@{}l*{3}{c}*{4}{c}@{}}
\toprule
\multirow{2}{*}{\textbf{Project}} & 
\multicolumn{3}{c}{\textbf{Python Features}} & 
\multicolumn{4}{c}{\textbf{C Features}} \\
\cmidrule(lr){2-4} \cmidrule(lr){5-8}
& \textbf{Decorators} & \textbf{First-Class Functions} & \textbf{Reflection} & 
\textbf{Function Pointers} & \textbf{Conditional Compilation} & \textbf{Macros} & \textbf{Goto} \\
\midrule
{\Aubio}          &         &         &         & \checkmark & \checkmark & \checkmark & \checkmark \\
\midrule
{\Bottleneck}     &         & \checkmark &         &         & \checkmark & \checkmark & \checkmark \\
\midrule
{\Bounter}        &         & \checkmark &         &         & \checkmark &         & \checkmark \\
\midrule
{\Cvxopt}         &         & \checkmark &         &         & \checkmark &         & \checkmark \\
\midrule
{\Japronto}       & \checkmark & \checkmark &         & \checkmark & \checkmark & \checkmark & \checkmark \\
\midrule
{\Libsmbios}      &         & \checkmark &         & \checkmark & \checkmark &         & \checkmark \\
\midrule
{\Pycurl}         & \checkmark & \checkmark &         &         & \checkmark &         & \checkmark \\
\midrule
{\Pygit}          &         & \checkmark &         &         &         &         & \checkmark \\
\midrule
{\Pytables}       &         & \checkmark &         &         & \checkmark &         & \checkmark \\
\midrule
{\Pyo}            &         &         &         & \checkmark & \checkmark & \checkmark &         \\
\midrule
{\Simplejson}     & \checkmark & \checkmark & \checkmark &         & \checkmark &         &         \\
\midrule
{\Tink}           &         & \checkmark &         &         &         &         &         \\
\midrule
{\Ultrajson}      &         &         &         &         & \checkmark &         &         \\
\midrule
\textbf{Total} & \textbf{21\%} & \textbf{71\%} & \textbf{7\%} & \textbf{29\%} & \textbf{79\%} & \textbf{29\%} & \textbf{64\%} \\
\bottomrule
\end{tabular}
}
\end{table}

\begin{table}[htbp]
\centering
\caption{Prevalence of targeted features in Java-C subjects}
\label{tab:java_prevalence}
\resizebox{0.5\textwidth}{!}{%
\begin{tabular}{@{}lcccccccc@{}}
\toprule
\multirow{2}{*}{\textbf{Project}} &
\multicolumn{4}{c}{\textbf{Java Features}} &
\multicolumn{4}{c}{\textbf{C Features}} \\
\cmidrule(lr){2-5} \cmidrule(lr){6-9}
& \textbf{Reflection} & \textbf{Lambda Expr.} & \textbf{Dynamic Proxy} & \textbf{Polymorphism}
& \textbf{Function Pointers} & \textbf{Conditional Compilation} & \textbf{Macro} & \textbf{Goto} \\
\midrule
\texttt{\Lwjgl}        & \checkmark & \checkmark &         & \checkmark & \checkmark & \checkmark & \checkmark & \checkmark \\
\midrule
\texttt{\Mdsplus}      & \checkmark &         &         & \checkmark & \checkmark & \checkmark & \checkmark &         \\
\midrule
\texttt{\Pijv}         & \checkmark & \checkmark &         & \checkmark &         & \checkmark &         &         \\
\midrule
\texttt{\PojavLauncher}& \checkmark & \checkmark &         & \checkmark &         &         &         & \checkmark \\
\midrule
\texttt{\Turbovnc}     & \checkmark &         & \checkmark & \checkmark & \checkmark & \checkmark & \checkmark & \checkmark \\
\midrule
\texttt{\Themis}       & \checkmark &         &         & \checkmark & \checkmark & \checkmark & \checkmark & \checkmark \\
\midrule
\texttt{\Espeakng}     & \checkmark &         &         & \checkmark & \checkmark & \checkmark & \checkmark & \checkmark \\
\midrule
\texttt{\Junixsocket}  & \checkmark &         &         & \checkmark & \checkmark & \checkmark & \checkmark &         \\
\midrule
\texttt{\Androidgifdrawable} & \checkmark &         &         & \checkmark & \checkmark & \checkmark & \checkmark &         \\
\midrule
\texttt{\ImBlockerFabric} & \checkmark & \checkmark &         & \checkmark & \checkmark &         &         &         \\
\midrule
\texttt{\Jep}          & \checkmark & \checkmark & \checkmark & \checkmark & \checkmark & \checkmark & \checkmark & \checkmark \\
\midrule
\texttt{\Hinn}         &         &         &         & \checkmark & \checkmark & \checkmark & \checkmark & \checkmark \\
\midrule
\texttt{\Termuxx}      & \checkmark & \checkmark &         & \checkmark & \checkmark & \checkmark & \checkmark & \checkmark \\
\midrule
\texttt{\Openjpeg}     &         &         &         & \checkmark & \checkmark & \checkmark & \checkmark &         \\
\midrule
\texttt{\Wolfssljni}   & \checkmark &         &         & \checkmark & \checkmark & \checkmark & \checkmark &         \\
\midrule
\textbf{Total (Java, n=15)} &
\textbf{87\%} & \textbf{40\%} & \textbf{13\%} & \textbf{100\%} &
\textbf{80\%} & \textbf{80\%} & \textbf{73\%} & \textbf{47\%} \\
\bottomrule
\end{tabular}%
}
\end{table}

Our analysis reveals that many of these targeted features are indeed utilized in practice, underscoring the need for analysis tools that can effectively handle them. Notably, a significant number of projects incorporate at least one, and often multiple, of these features, indicating that complex language constructs are not mere edge cases but are actively employed in real-world multilingual software development.

Among the Python features analyzed, First-Class Functions were observed with high frequency, present in 10 out of the 14 projects. Decorators were also utilized in three projects (\texttt{\small japronto}, \texttt{\small pycurl}, \texttt{\small simplejson}), indicating their adoption for modifying or extending behavior. In contrast, specific dynamic features like Reflection (found in \texttt{\small simplejson}) and Dynamic Imports (found in \texttt{\small pyo}) were less common in this particular dataset, each appearing in only one project. Lambda Functions, as per our analysis focusing on the features presented in the results table, were not detected in any analyzed project.

For the Java features in Java-C subjects, reflection is pervasive: Reflection appears in 87\% of projects, indicating that runtime type inspection and reflective invocation are routine in Java-native integration. Lambda expressions are also common (40\%), often compiling into synthetic classes/methods and complicating call-graph recovery for callback-style code. Dynamic proxies, while less frequent (13\%), add another layer of runtime indirection via invocation handlers. Polymorphism (i.e., dynamic dispatch on member methods) occurs in all projects, showing that native interaction is typically encapsulated within Object-Oriented APIs rather than standalone static entrypoints. Overall, Java-C systems frequently combine reflection, functional abstractions, and proxy-based dispatch, motivating analyses that can conservatively model these dynamic behaviors for cross-language reasoning.

For the C features in Python-C subjects, Conditional Compilation is prevalent, reflecting heavy reliance on platform- and build-dependent code paths. Setjmp/Longjmp is also common, while Function Pointers appear in 4 projects and Goto in 3, both introducing control-flow indirection. Macro-related constructs were not observed in this Python-C prevalence analysis. In contrast, Java-C subjects exhibit denser native-side complexity: Function Pointers and Conditional Compilation each appear in 80\% of projects, Macro Calls in 73\%, and Goto in 47\%, suggesting that preprocessor-driven variation and indirect control flow are more typical in Java-C settings.

In both Python-C and Java-C subjects, we observe that “hard” language feature usages are routine rather than exceptional. Python-C projects frequently rely on first-class functions and non-trivial native control variability (e.g., conditional compilation and low-level control-flow constructs), while Java-C projects consistently embed native interaction inside object-oriented APIs and commonly introduce additional indirection via reflection and lambda-based callbacks. On the native side, Java-C subjects exhibit denser preprocessor- and indirection-heavy patterns (e.g., function pointers and macros), further complicating static cross-language analysis.

% These findings confirm that challenging language features, which can obscure information flow and introduce vulnerabilities, are present across a range of real-world Python/C projects, which reinforces the importance of developing sophisticated static analysis techniques, such as those in {\tech}, capable of accurately interpreting these constructs in a cross-language setting.

\find{
%Our prevalence results show that 
Feature-induced analysis challenges are common in real-world cross-language systems, motivating {\tech} that can soundly model such constructs to avoid missed or spurious information flows.
}

\subsection{RQ2: Effectiveness of {\tech}}

\textbf{Expert Panel Accuracy.} The expert panel is the core component responsible for interpreting complex language features that traditional static analyzers fail. To evaluate its performance, we assessed its ability to identify implicit control and data flows. The panel achieved a precision of 0.979, a recall of 0.873, and an F1-score of 0.918 based on xFlowBench. These results demonstrate the expert panel's high reliability. 

\textbf{Source and Sink Identification.} 
Table~\ref{tab:source_sink_precision_merged} reports per-project precision and volume of the identified sources/sinks for Python-C and Java-C subjects, respectively. For sources, precision is consistently moderate-to-high, averaging 80\% in Python-C and 83\% in Java-C, with substantial cross-project variation in counts (Avg.\ 305 sources/project for Python-C vs.\ 140 for Java-C). Larger projects naturally contribute  more candidates (e.g., \texttt{\small pytables}: 909 sources; \texttt{\small lwjgl}: 566 sources), confirming that source identification dominates the workload in both settings.
For sinks, precision is generally higher and more stable, 
% averaging 95\% in Python-C and 93\% in Java-C, 
while sink counts remain much smaller, indicating that security-relevant endpoints are relatively sparse compared to input-like sources. Overall, on most of the subject projects {\tech} achieved high sink precision across both language pairs.
% We evaluated {\tech}'s LLM-driven ability to automatically identify security-sensitive sources and sinks on 14 real-world projects. As shown in Table~\ref{tab:source_sink_precision} and Table~\ref{tab:source_sink_precision_java}, {\tech} demonstrates high precision in this task. The superior performance on sinks is likely because sink functions (e.g., memory allocation) are often more standardized and security-critical, thus easier for the LLM to recognize. In contrast, data sources can be more varied and application-specific, leading to slightly lower, yet still effective, precision.

% Notably, the performance varies across projects, with source precision being very high for some (e.g., 88\% for \texttt{\small japronto}) but lower for others (e.g., 16\% for \texttt{\small tink}). This variance suggests that project-specific coding conventions and complexity, especially in specialized libraries like the cryptographic tink, can influence the accuracy of LLM-based detection.

%\section{Additional Evaluation and Results}\label{sec:moreevalresults}
%\subsection{Source/Sink Identification}\label{ssec:srcsinkresults}
%Table~\ref{tab:source_sink_precision_merged} shows the detailed evaluation results of the automated source/sink identification component in {\tech}. We have three main observations. 

%Three patterns stand out. 
More specifically, the results reveal three notable findings. 
\emph{First}, sink identification is overall more reliable than source identification: sink precision averages 95\% (Python--C) and 93\% (Java--C), with most projects at or above 90\%, whereas source precision averages 80\% and 83\% respectively. This asymmetry reflects the underlying task: sinks are syntactically narrow (a small set of dangerous APIs and write/log/exec sites), while sources span a much broader and more heterogeneous surface (user inputs, deserialized data, environment, IPC, file/network reads), inviting more borderline judgments. \emph{Second}, sources outnumber sinks by roughly an order of magnitude in both settings (Python--C: 305 vs.\ 32 on average; Java--C: 140 vs.\ 16), confirming that source enumeration dominates the candidate workload and that downstream slicing must scale to many sources per sink rather than the reverse. \emph{Third}, project-to-project variation in counts is indicated mainly by codebase size and API surface (e.g., pytables: 909 sources; lwjgl: 566), but precision is largely \emph{insensitive} to scale --- both source and sink precision stay within a tight band across small (bounter, imblockerfabric) and large projects --- indicating that the LLM-based source/sink identifier %degrades gracefully rather than collapsing 
does not collapse 
on bigger or more idiomatically diverse codebases.

\textbf{Data Flow Path Detection.} We evaluated {\tech} against xLoc, a state-of-the-art learning-based baseline, on the task of identifying valid, exploitable data-flow paths from sources to sinks. As shown in Table~\ref{tab:detection_performance}, {\tech} successfully identified 454 true positive cross-language data-flow paths across the evaluated projects, achieving an overall precision of 76.8\%. It even achieved perfect 100\% precision on five projects, including \texttt{\small bounter} and \texttt{\small cvxopt}.

Most significantly, xLoc failed to detect any vulnerability path. This result highlights the importance of explicitly modeling the challenging language features that mediate cross-language source-to-sink flows. When these features are not captured, cross-language reachability remains incomplete and the corresponding paths cannot be recovered. In contrast, {\tech}’s feature-aware augmentation recovers the missing cross-language control/data-flow facts before propagation, making these otherwise disconnected information flow paths visible. 

\begin{table}[tp]
\centering
    \caption{Description of ablated versions}
    \label{tab:ablated-new}
    \resizebox{.9\linewidth}{!}{%
    \begin{tabular}{l|c|c|c}
    \toprule
         & Augmented Control Flow & Feature-Oriented Data Flow & Expert Panel \\
    \midrule
    (1) w/o FDF & \ding{51} & \ding{55} &  \ding{51}\\
    \midrule
    (2) w/o ACF   & \ding{55} & \ding{51} & \ding{51} \\
    \midrule
    (3) w/o EP & \ding{51} & \ding{51} & \ding{55} \\
    \bottomrule
    \end{tabular}
    }
\end{table}

\find{{\tech} is effective end-to-end for cross-language vulnerability detection: it reliably identifies sources/sinks and recovers accurate source-to-sink data-flow paths at scale across both Python-C and Java-C projects.}

% \subsection{RQ3: Does each of its components contribute significantly to {\tech}'s performance?}
\subsection{RQ3: Ablation Studies}

% To understand the individual contributions of the core components within {\tech}, we conduct an ablation study. 
We evaluate the impact of the \textbf{Augmented Control Flow (ACF)} analysis, which enriches the program's graph representation to reveal hidden execution paths; the \textbf{Feature-Oriented Data Flow (FDF)} reasoning, which tracks data propagation with a focus on implicit flows; and the \textbf{Expert Panel (EP)}, which integrates LLMs for advanced reasoning and refinement. We created three ablated versions of {\tech} by selectively disabling each component, as detailed in Table~\ref{tab:ablated-new}. By comparing the effectiveness of these versions against the full {\tech} design (which found 272 true positives at 76.8\% precision) in detecting cross-language vulnerabilities, we can quantify the contribution and importance of each component.

\begin{table}[t]

\caption{Results of ablation studies}
\label{tab:ablation_analysis}
\centering
\resizebox{0.9\columnwidth}{!}{%
\begin{tabular}{l|rrr|rrr|rrr}
\hline
\multirow{2}{*}{\textbf{Project}} & \multicolumn{3}{c|}{\textbf{Ablation 1}} & \multicolumn{3}{c|}{\textbf{Ablation 2}} & \multicolumn{3}{c}{\textbf{Ablation 3}} \\
\cline{2-4} \cline{5-7} \cline{8-10}
 & \textbf{\#Paths} & \textbf{\#TPs} & \textbf{Prec} & \textbf{\#Paths} & \textbf{\#TPs} & \textbf{Prec} & \textbf{\#Paths} & \textbf{\#TPs} & \textbf{Prec} \\
\hline
\Aubio & 37 & 27 & 73\% & 21 & 21 & 100\% & 21 & 21 & 100\% \\
\Bottleneck & 62 & 22 & 35\% & 17 & 15 & 88\% & 19 & 16 & 84\% \\
\Bounter & 66 & 28 & 42\% & 20 & 20 & 100\% & 26 & 20 & 77\% \\
\Cvxopt & 3 & 1 & 33\% & 0 & 0 & - & 0 & 0 & - \\
\Japronto & 48 & 21 & 44\% & 1 & 1 & 100\% & 13 & 13 & 100\% \\
\Libsmbios & 102 & 31 & 30\% & 47 & 26 & 55\% & 48 & 26 & 54\% \\
\Msgpack & 11 & 3 & 27\% & 3 & 3 & 100\% & 3 & 3 & 100\% \\
\Pycurl & 49 & 3 & 6\% & 4 & 3 & 75\% & 4 & 3 & 75\% \\
\Pygit & 83 & 35 & 42\% & 34 & 30 & 88\% & 39 & 34 & 87\% \\
\Pyo & 57 & 0 & 0\% & 0 & 0 & - & 0 & 0 & - \\
\Pytables & 49 & 0 & 0\% & 0 & 0 & - & 0 & 0 & - \\
\Simplejson & 15 & 24 & 160\% & 22 & 22 & 100\% & 22 & 22 & 100\% \\
\Tink & 163 & 65 & 40\% & 81 & 58 & 72\% & 87 & 59 & 68\% \\
\Ultrajson & 41 & 12 & 29\% & 10 & 10 & 100\% & 12 & 11 & 92\% \\
\hline
\Lwjgl & 195 & 54 & 28\% & 49 & 38 & 78\% & 66 & 45 & 68\% \\
\Mdsplus & 24 & 7 & 29\% & 7 & 6 & 86\% & 8 & 5 & 63\% \\
\Pijv & 20 & 6 & 30\% & 6 & 5 & 83\% & 7 & 5 & 71\% \\
\PojavLauncher & 30 & 7 & 23\% & 8 & 6 & 75\% & 9 & 5 & 56\% \\
\Turbovnc & 22 & 6 & 27\% & 6 & 5 & 83\% & 8 & 5 & 63\% \\
\Themis & 40 & 11 & 28\% & 11 & 9 & 82\% & 14 & 9 & 64\% \\
\Espeakng & 13 & 3 & 23\% & 2 & 2 & 100\% & 4 & 2 & 50\% \\
\Junixsocket & 45 & 15 & 33\% & 14 & 12 & 86\% & 15 & 12 & 80\% \\
\Androidgifdrawable & 28 & 7 & 25\% & 7 & 6 & 86\% & 8 & 5 & 63\% \\
\ImBlockerFabric & 0 & 0 & - & 0 & 0 & - & 0 & 0 & - \\
\Jep & 90 & 26 & 29\% & 25 & 19 & 76\% & 30 & 22 & 73\% \\
\Hinn & 15 & 3 & 20\% & 2 & 2 & 100\% & 4 & 2 & 50\% \\
\Termuxx & 48 & 15 & 31\% & 15 & 12 & 80\% & 16 & 12 & 75\% \\
\Openjpeg & 38 & 12 & 32\% & 11 & 9 & 82\% & 13 & 10 & 77\% \\
\Wolfssljni & 35 & 10 & 29\% & 10 & 8 & 80\% & 12 & 8 & 67\% \\
\hline
\textbf{Total} & \textbf{1429} & \textbf{454} & \textbf{31.8\%} & \textbf{433} & \textbf{348} & \textbf{80.4\%} & \textbf{508} & \textbf{375} & \textbf{73.8\%} \\
\hline
\end{tabular}%
}
%\vspace{-16pt}
\end{table}

\begin{table*}[t]
\caption{{\tech} cost breakdown}
\label{tab:performance_analysis}
\centering
\resizebox{\textwidth}{!}{%
\begin{tabular}{ll|*{15}{c}|*{16}{c}}
\toprule
& 
& {\Aubio}
& {\Bottleneck}
& {\Bounter}
& {\Cvxopt}
& {\Japronto}
& {\Libsmbios}
& {\Msgpack}
& {\Pycurl}
& {\Pygit}
& {\Pyo}
& {\Pytables}
& {\Simplejson}
& {\Tink}
& {\Ultrajson}
& \textbf{Average}
& {\Lwjgl}
& {\Mdsplus}
& {\Pijv}
& {\PojavLauncher}
& {\Turbovnc}
& {\Themis}
& {\Espeakng}
& {\Junixsocket}
& {\Androidgifdrawable}
& {\ImBlockerFabric}
& {\Jep}
& {\Hinn}
& {\Termuxx}
& {\Openjpeg}
& {\Wolfssljni}
& \textbf{Average} \\
\midrule
\multirow{3}{*}{\textbf{Phase 1}}
& \textbf{Step 1 Time (h)}
& 0.36 & 0.21 & 0.58 & 0.36 & 0.35 & 0.06 & 0.08 & 0.11 & 0.08 & 0.13 & 1.23 & 0.14 & 0.41 & 0.78 & 0.35
& 0.45 & 0.39 & 0.18 & 0.20 & 0.31 & 0.25 & 0.20 & 0.20 & 0.14 & 0.12 & 0.16 & 0.18 & 0.14 & 0.28 & 0.21 & 0.23 \\
& \textbf{Step 2 Time (h)}
& 3.63 & 1.82 & 0.67 & 4.13 & 2.58 & 1.13 & 0.62 & 3.37 & 2.42 & 11.26 & 7.19 & 0.99 & 4.78 & 0.31 & 3.21
& 10.51 & 5.37 & 2.76 & 3.46 & 4.54 & 4.34 & 2.97 & 3.77 & 2.05 & 0.99 & 4.16 & 2.62 & 2.80 & 4.56 & 3.60 & 3.90 \\
& \textbf{Step 2 Tokens (M)}
& 2.80 & 0.98 & 0.49 & 4.61 & 1.35 & 0.84 & 0.41 & 1.94 & 2.40 & 15.00 & 8.32 & 0.69 & 3.63 & 0.32 & 3.13
& 6.02 & 1.66 & 1.04 & 1.28 & 1.30 & 1.56 & 0.89 & 1.62 & 1.18 & 0.59 & 2.54 & 0.93 & 1.63 & 1.55 & 1.40 & 1.68 \\
\midrule
\multirow{4}{*}{\textbf{Phase 2}}
& \textbf{Step 1 Time (h)}
& 4.60 & 2.56 & 0.44 & 4.15 & 1.48 & 1.32 & 6.13 & 0.50 & 1.11 & 8.10 & 3.63 & 6.88 & 6.17 & 7.87 & 3.93
& 10.51 & 5.37 & 2.76 & 3.46 & 4.54 & 4.34 & 2.97 & 3.77 & 2.05 & 0.99 & 4.16 & 2.62 & 2.80 & 4.56 & 3.60 & 3.90 \\
& \textbf{Step 2 Time (h)}
& 4.48 & 4.39 & 2.61 & 7.36 & 4.83 & 4.18 & 5.58 & 0.63 & 4.72 & 9.45 & 4.37 & 5.47 & 8.24 & 5.36 & 5.13
& 14.01 & 7.17 & 3.68 & 4.62 & 6.05 & 5.78 & 3.96 & 5.03 & 2.73 & 1.32 & 5.55 & 3.49 & 3.73 & 6.07 & 4.79 & 5.20 \\
& \textbf{Step 2 Tokens (M)}
& 0.12 & 0.15 & 0.15 & 0.90 & 0.12 & 0.24 & 0.11 & 0.33 & 0.28 & 2.10 & 2.20 & 0.14 & 0.62 & 0.09 & 0.54
& 0.90 & 0.18 & 0.16 & 0.21 & 0.17 & 0.25 & 0.13 & 0.26 & 0.20 & 0.11 & 0.45 & 0.14 & 0.28 & 0.24 & 0.22 & 0.26 \\
& \textbf{Avg Time per Query (s)}
& 2.50 & 3.18 & 1.62 & 5.70 & 2.66 & 1.85 & 1.85 & 1.81 & 1.84 & 4.23 & 2.20 & 1.87 & 4.86 & 1.68 & 2.70
& 4.11 & 2.80 & 2.25 & 2.42 & 2.63 & 2.61 & 2.29 & 2.51 & 2.04 & 1.62 & 2.63 & 2.21 & 2.26 & 2.66 & 2.45 & 2.50 \\
\midrule
\multirow{1}{*}{\textbf{Resource}}
& \textbf{Memory Peak (MB)}
& 274 & 209 & 167 & 338 & 213 & 183 & 183 & 230 & 224 & 2009 & 606 & 185 & 315 & 187 & 380
& 1672 & 857 & 408 & 483 & 654 & 594 & 422 & 526 & 359 & 236 & 601 & 388 & 444 & 632 & 502 & 585 \\
\bottomrule
\end{tabular}%
}
\end{table*}

The results in Table~\ref{tab:ablation_analysis} reveal the distinct and synergistic roles of each component. Disabling the \textbf{FDF} analysis (Ablation 1) dramatically degrades precision. 
% While this version successfully identifies all 454 true positive paths, it does so at the cost of being flooded by false positives, causing the overall precision to plummet to 31.8\% from an explosion in reported paths. 
This demonstrates that FDF is not essential for reachability but is crucial for pruning infeasible paths by verifying actual data propagation, thereby eliminating a significant number of false alarms.
Additionally, removing the \textbf{ACF} analysis (Ablation 2) severely impacts recall. The number of true positives found drops by nearly 23\%, as many valid cross-language execution paths are never discovered. Without ACF's ability to resolve dynamic and implicit control flows (e.g., via function pointers or decorators), the foundational graph representation is incomplete, rendering entire classes of vulnerabilities invisible to the subsequent data-flow analysis. 

% The precision, however, remains high (80.4\%) for the paths it can analyze, as the FDF component is still effective on the limited graph.

Finally, operating without the \textbf{EP} (Ablation~3) degrades both precision and recall. Compared to Ablation~2, precision drops from 80.4\% to \textbf{73.8\%} and true positives increase only modestly (from 348 to \textbf{375}), indicating that EP contributes in two complementary ways: it improves recall by correctly interpreting complex language features that the symbolic engine alone cannot resolve, and it improves precision by providing context-aware reasoning to validate or refute borderline taint flows that would otherwise be misclassified.

% In summary, the experimental results highlight that each component is vital. ACF provides the necessary foundational path coverage, FDF delivers the precision needed to make the results usable, and the EP provides the deep reasoning required to handle modern language complexities. Together, these components form a synergistic framework that is significantly more effective than any of its parts in isolation.

% \find{ACF is critical for recall: without it, unmodeled implicit control flows break cross-language reachability and many true vulnerability paths are never discovered. FDF is critical for precision: without it, path enumeration is flooded by infeasible flows; EP further improves both by resolving hard language features and validating borderline taint flows beyond the symbolic engine.}
\find{Each component of {\tech} is vital. ACF provides necessary foundational path coverage, FDF delivers the precision needed to make the results usable, and the EP provides the deep reasoning required to handle modern language complexities.}

% \subsection{RQ4: How efficient is {\tech} on real-world systems?}
\subsection{RQ4: Efficiency of {\tech}}\label{RQ4}

To comprehensively evaluate the runtime efficiency of {\tech}, we conducted experiments on 29 real-world benchmarks. Our evaluation focused on key performance metrics: total analysis time, LLM token consumption, and peak memory usage, contextualized by the lines of code (LOC) of each project. 
% Figure~\ref{fig:efficiency} summarizes the overall performance, while 
Table~\ref{tab:performance_analysis} provides a breakdown of resource consumption across {\tech}'s two main analysis phases.

Table~\ref{tab:performance_analysis} shows that {\tech}'s runtime is dominated by the two LLM-centric steps: Phase~1 Step~2 (CICFG augmentation) and Phase~2 Step~2 (Instructed Taint Refinement). These steps account for most of the wall-clock time and nearly all token consumption, as they offload the reasoning over hard-to-model cross-language constructs (e.g., function pointers, JNI boundary semantics) to the expert panel. In contrast, the non-LLM steps (e.g., CICFG construction and taint approximation in Phase~1 Step~1, and the initial analysis stage in Phase~2 Step~1) are comparatively lightweight.

Across Python-C subjects, {\tech}'s runtime cost averages 3.21h and 5.13h for two LLM steps (Phase~1 Step~2 and Phase~2 Step~2), consuming 3.13M and 0.54M tokens, respectively. For Java-C subjects, the same steps average 3.90h and 5.20h, with lower token usage (1.68M and 0.26M), indicating that time cost is driven primarily by LLM reasoning overhead rather than tokens alone. The per-query latency remains stable and practical in both groups (2.70s on Python-C vs. 2.50s on Java-C).

Peak memory usage scales moderately and remains practical for commodity workstations. The Python-C subjects peak at 380MB on average, with most projects well under 1GB and a single outlier (\texttt{\small pyo}) reaching $\sim$2GB. The Java-C subjects exhibit a higher average peak of 585MB, with \texttt{\small lwjgl3} as an outlier at 1672MB. Overall, these footprints demonstrate that {\tech}'s memory requirements are manageable in real-world cross-language analysis.

% Overall, our evaluation demonstrates that {\tech} achieves a reasonable balance between analytical depth and resource efficiency. While the analysis of large, complex systems is computationally intensive, the costs are primarily driven by the necessary and targeted application of LLM-based reasoning to overcome the fundamental limitations of traditional static analysis. The framework's architecture effectively contains these costs, particularly memory usage, making it a viable and scalable solution for discovering complex vulnerabilities in real-world, multi-language software.

\find{{\tech} is efficient, with runtime and token cost dominated by the CICFG augmentation and instructed taint refinement, while per-query latency stays practical. Peak memory usage is moderate, indicating {\tech} is scalable for real-world cross-language analysis.}

% \subsection{RQ5: Can {\tech} find cross-language vulnerabilities?}\label{RQ4}
\subsection{RQ5: Real-world Bug Discovery}
\newcommand{\tinycircnum}[1]{%
\tikz[baseline=(char.base)]{
    \node[
      shape=circle,
      fill=black,
      inner sep=0pt,
      minimum size=0.7em,
      text height=0.5em,  % Controls vertical centering
      text depth=0.1em,  % Controls baseline offset
      align=center
    ] (char) {\sffamily\bfseries\color{white}\scriptsize #1};
  }%
}
% \begin{table}[t]
% \caption{Comparison of Detection Tools by Project}
% \label{tab:vul_label}
% \centering

% \begin{tabular}{l|cccc}
% \hline
% \textbf{Subject} & \textbf{PolyFlow} & \textbf{MultiQL} & \textbf{xLoc} \\
% \hline
% \Aubio & 1 & 0 & 0 \\
% \Bounter & 3 & 0 & 0 \\
% \Japronto & 2 & 0 & 0 \\
% \Pygit & 1 & 0 & 0 \\
% \Cvxopt & 1 & 0 & 0 \\
% \hline
% \textbf{Total} & \textbf{8} & \textbf{0} & \textbf{0} \\
% \hline
% \end{tabular}

% \end{table}

% \begin{wraptable}{l}{0.45\linewidth}
%     \caption{Comparison of detection tools by project}
%     \label{tab:vul_label}
%     \centering
%     % 表格的tabular部分保持不变
%     \scalebox{0.71}{
%         \begin{tabular}{l|cccc}
%         \hline
%         \textbf{Subject} & \textbf{{\tech}} & \textbf{MultiQL} & \textbf{xLoc} \\
%         \hline
%         \Aubio & 1 & 0 & 0 \\
%         \Bounter & 2 & 0 & 0 \\
%         \Ultrajson & 1 & 0 & 0 \\
%         \Japronto & 2 & 0 & 0 \\
%         \Pygit & 1 & 0 & 0 \\
%         \Cvxopt & 1 & 0 & 0 \\
%         \hline
%         \textbf{Total} & \textbf{8} & \textbf{0} & \textbf{0} \\
%         \hline
%         \end{tabular}
%     }
% \end{wraptable}

\begin{table}[t]
%\vspace{-4pt}
\centering
\caption{Summary of real-world bug discovery}
\label{tab:vul_label}
\resizebox{\columnwidth}{!}{%
\begin{tabular}{lcccccccccc}
\toprule
 & \textbf{Bounter} & \textbf{Pygit2} & \textbf{Aubio} & \textbf{Ultrajson} & \textbf{Japronto} & \textbf{Bitarray} & \textbf{Cvxopt} & \textbf{Lwjgl} & \textbf{Jep} & \textbf{Total} \\
\midrule
\textbf{\#Bugs Found} & 2 & 1 & 1 & 1 & 2 & 4 & 1 & 3 & 2 & 17 \\
\bottomrule
\end{tabular}%
}
%\vspace{-4pt}
\end{table}

\begin{figure}[tp]
    \centering
    \includegraphics[width=1\linewidth]{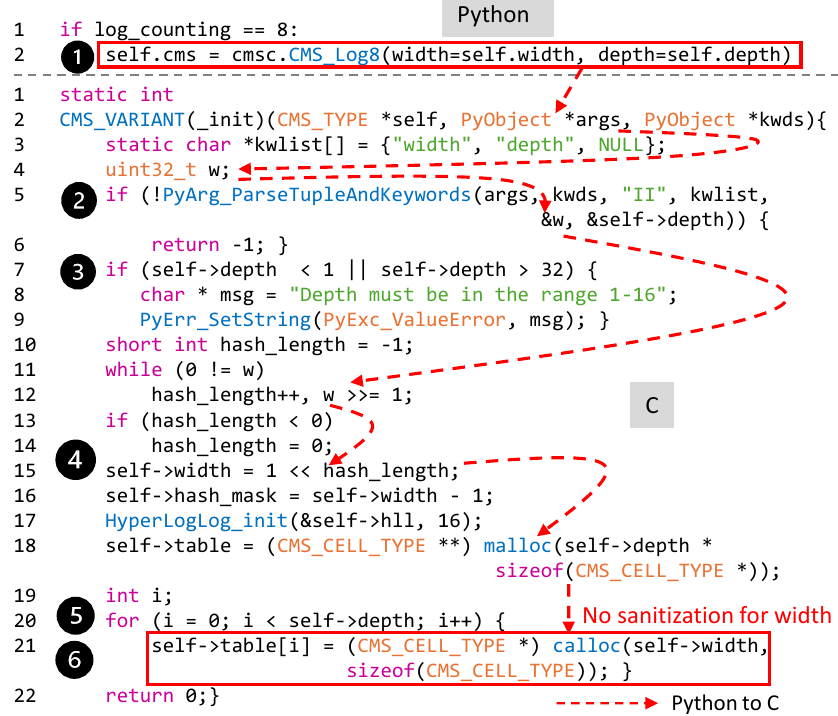}
    \caption{A vulnerability case found in Python-C project.}
    \label{fig:bounter}
    \vspace{2pt}
\end{figure}

Table~\ref{tab:vul_label} summarizes the number of bugs detected per benchmark, along with the total count. Notably, one of the reported issues has already been assigned a CVE identifier (CVE-2025-55562). Also, another bug has been confirmed by the maintainers and fixed via an upstream pull request~\footnote{ \url{https://github.com/LedFx/aubio-ledfx/pull/22}}. 
%Detailed per-bug reports are provided in Appendix~\ref{sec:real-world-bug}. 

%\section{Details on detected real-world bugs} \label{sec:real-world-bug}
% requires: \usepackage{booktabs}
% optional but recommended: \usepackage{tabularx}

% requires: \usepackage{booktabs}

% requires: \usepackage{booktabs}

\begin{table*}[t]
\centering
\caption{Details on the real-world cross-language bugs found by {\tech} in Python--C and Java--C systems, with assigned CVE identifiers and reporting status (the specific CVE numbers are obviated for ethical considerations)}
\label{tab:crosslang_bugs_status}
\footnotesize
\setlength{\tabcolsep}{4pt}
\resizebox{1\textwidth}{!}{%
\begin{tabular}{r l p{0.560\textwidth} l l l}
\toprule
\textbf{Bug ID} & \textbf{Bug Type} & \textbf{Bug Description} & \textbf{Project} & \textbf{CVE} & \textbf{Reporting Status} \\
\midrule
1  & Crash / NULL dereference (DoS) &
Huge (but valid) CountMinSketch width/depth can make \texttt{calloc()} fail; the NULL table is later dereferenced (e.g., \texttt{self->table[i][bucket]}), crashing the process. &
bounter & --- & pending \\

2  & Crash / NULL dereference (DoS) &
\texttt{CMS\_VARIANT(\_init)} allocates internal tables based on user width/depth; allocation failure under extreme parameters is not checked and later triggers NULL dereference. &
bounter & \ding{51} & confirmed \\

3  & Heap out-of-bounds read (deserialization) &
\texttt{HT\_Basic.\_\_setstate\_\_} \texttt{memcpy}s attacker-controlled pickle bytearrays into \texttt{self->table}, \texttt{self->histo}, and \texttt{self->hll.registers} without checking the source byte-length, so an undersized bytearray causes reads past the buffer. &
bounter & --- & pending \\

4  & Buffer overflow (type confusion in size output) &
\texttt{pgit\_odb\_backend\_read(*)} passes \texttt{size\_t*} into \texttt{PyArg\_ParseTuple(...,"ny\#"...)} causing size corruption; the resulting length is used in alloc/copy (\texttt{...data\_alloc} $\rightarrow$ \texttt{memcpy}), enabling overflow. &
pygit2 & \ding{51} & confirmed \\

5  & Buffer handling bug  &
\texttt{aubio\_sampler\_load} copies a \texttt{PATH\_MAX}-bounded URI via \texttt{strncpy} without guaranteeing NUL-termination, which can cause later string operations to read past the buffer. &
aubio & \ding{51} & confirmed \\

6  & Out-of-bounds read (buffer over-read) &
In \texttt{Buffer\_EscapeStringValidated}, the 4-byte UTF-8 path may \texttt{memcpy} a \texttt{JSUTF32} when fewer than 4 bytes remain, reading past the input buffer. &
ultrajson & \ding{51} & confirmed \\

7  & Resource exhaustion (DoS) &
\texttt{Matcher\_init} derives \texttt{buffer\_len} from user-influenced route compilation without bounds checks, then \texttt{malloc(buffer\_len)} and \texttt{memcpy} can force excessive allocation (OOM/DoS). &
Japronto & \ding{51} & confirmed \\

8  & Heap-based buffer overflow &
The HTTP ``gather'' logic allocates a fixed max buffer but copies unbounded request bodies/chunks; when \texttt{len > GATHER\_MAX\_LEN}, the copy overflows the heap buffer. &
Japronto & \ding{51} & fixed \\

9  & Resource Exhaustion / Uncontrolled Memory Allocation &
Pickle-controlled \texttt{nbits} flows into \texttt{bitarray\_tolist}, which calls \texttt{PyList\_New(self->nbits)} with no upper bound, enabling massive allocation (OOM/DoS). &
bitarray & --- & pending \\

10  & Resource Exhaustion / Uncontrolled Memory Allocation &
Pickle-controlled \texttt{nbits} flows into \texttt{bitarray\_unpack}, which allocates \texttt{PyBytes\_FromStringAndSize(NULL, self->nbits)} without a cap, enabling one-shot huge allocation (OOM/DoS). &
bitarray & --- & pending \\

11  & Resource Exhaustion / Uncontrolled Memory Allocation  &
\texttt{ba2hex} computes \texttt{out\_len = 2 * BYTES(nbits)} from pickle-controlled \texttt{nbits} and allocates output bytes without bounds, allowing GB-scale allocation (OOM/DoS). &
bitarray & --- & pending \\

12  & Resource exhaustion  &
\texttt{to01()} allocates a temporary buffer of \texttt{self->nbits} bytes via \texttt{PyMem\_Malloc} and builds a Python string; with attacker-controlled \texttt{nbits}, this enables OOM/DoS. &
bitarray & \ding{51} & confirmed \\

13  & Format-string vulnerability &
User input reaches \texttt{validate(char*)} and is dispatched via an indirect call \texttt{validator(input)}; if the callee is \texttt{printf}-like, attacker-controlled format strings can trigger a format-string bug. &
cvxopt & --- & pending \\

\midrule
14  & Unsafe indirect call  &
LWJGL JNI wrapper casts Java-controlled \texttt{\_\_functionAddress} to a function pointer and invokes it without validation, enabling arbitrary native calls or crashes. &
LWJGL & --- & pending \\

15  & Arbitrary memory write  &
LWJGL \texttt{nputDouble} casts Java-controlled \texttt{ptrAddress} to \texttt{double*} and writes through it without checks, enabling arbitrary write and memory corruption. &
LWJGL & --- & pending \\

16  & Unchecked length copy   &
LWJGL \texttt{nmemmove} casts Java-provided addresses to pointers and calls \texttt{memmove(..., (size\_t)count)} with an untrusted \texttt{count}, enabling out-of-bounds read/write. &
LWJGL & --- & pending \\

17  & Stateful code injection / unexpected execution &
Jep interactive mode buffers non-compilable lines in \texttt{evalLines}; a later flush concatenates and evaluates them, enabling cross-request injection if an instance is reused across trust boundaries. &
Jep & --- & pending \\

18  & Crash / NULL dereference  &
Under OOM, \texttt{GetStringUTFChars} may return NULL; Jep forwards this pointer into \texttt{pyembed\_eval} without a NULL check, causing a native crash. &
Jep & --- & pending \\
\bottomrule
\end{tabular}%
}
\end{table*}

More specifically, Table~\ref{tab:crosslang_bugs_status} summarizes the full set of cross-language bugs identified in our study and standardizes them into a consistent reporting view. Each row corresponds to one bug and records sixfive fields: a unique Bug ID, a normalized Bug Type (e.g., NULL dereference, OOB read, heap overflow, resource exhaustion), a concise Bug Description, the affected Project, the assigned CVE identifier (if any), and the current Reporting Status. Status labels indicate the lifecycle stage of each finding: pending (not yet confirmed upstream or not yet resolved), 
confirmed (validated by developer or via CVE assignment), 
and fixed (patched by maintainers). 
%Overall, the table serves as an appendix-grade index that links projects to concrete low-level failure modes and their disclosure/patch progress in a single place.

We select one previously unknown bug discovered by {\tech} to conduct a detailed case study. \texttt{Bounter}
% ~\footnote{\url{https://github.com/piskvorky/bounter}} 
is a Python-C library designed for high-speed, memory-bounded frequency counting in massive datasets. The vulnerability identified by {\tech} exists in the C implementation of the Count-Min Sketch (CMS) initialization, specifically the \texttt{\small CMS\_VARIANT(\_init)} initializer in \texttt{\small cbounter/cms\_common.c}. 
Notably, this project uses a \emph{macro-generated variant scheme}: the Python-visible constructor name (e.g., \texttt{\small CMS\_Log8} at \tinycircnum{1}) and the underlying C initializer entry are not written with the same identifier in source code, because the concrete entry name is produced/aliased through macro expansion (e.g., a variant-specific \texttt{\small *\_init}).
Therefore, without handling our \textbf{C macro feature}, the Python to C transition would appear disconnected (missing call edge).

As shown in Figure~\ref{fig:bounter}, the user-provided width and depth parameters flow from Python into the C initialization function and are parsed using \texttt{\small PyArg\_ParseTupleAndKeywords} (\tinycircnum{2}). 
Linking this Python-level constructor to the correct native initializer is non-trivial because the C entry is macro-generated/aliased (written as \texttt{\small CMS\_VARIANT(\_init)} in the source). 
By resolving this macro-induced name mismatch, {\tech} restores the missing cross-language call edge, after which the user-provided width and depth are parsed in the C initializer via \texttt{\small PyArg\_ParseTupleAndKeywords} (\tinycircnum{2}). 
The code subsequently sanitizes the depth parameter by enforcing a strict range check $[1, 32]$ (\tinycircnum{3}), ensuring it does not exceed a small, fixed upper bound. However, the width parameter lacks equivalent sanitization. It is processed to calculate \texttt{\small self->width} by rounding the input down to the nearest power of two (\tinycircnum{4}), but no check prevents this value from becoming excessively large based on the input.

The function then proceeds to allocate memory for each row of the \texttt{\small CMS} table within a loop (\tinycircnum{5}). Inside this loop, the calloc function is invoked (\tinycircnum{6}): \texttt{\small{ self->table[i] = (CMS\_CELL\_TYPE *) calloc(self->width, sizeof(CMS\_CELL\_TYPE));}}. From a static analysis perspective focusing on input validation, the lack of strict sanitization for the width parameter is critical. A very large user input for width can lead to a large \texttt{\small{self->width}} value, causing the calloc at line 31 to request a potentially huge amount of memory. This significantly increases the risk of \texttt{\small calloc} failing and returning NULL. Crucially, the return value from \texttt{\small calloc} is assigned directly to \texttt{\small self->table[i]} without any check for NULL. Should calloc fail due to large size requested (stemming from the insufficiently sanitized width), \texttt{\small self->table[i]} becomes NULL. 
% While this assignment itself does not crash the program, it introduces a null pointer into the CMS structure, setting the stage for a null-pointer dereference vulnerability when 
% Subsequent operations (e.g., increment, get) attempt to access this specific table row \texttt{\small self->table[i]} expecting valid memory. 
Thus, the vulnerability originates from the unsanitized width input leading to a potential \texttt{\small calloc} failure whose result is used unsafely.

\begin{figure}[tp]
    \centering
    \includegraphics[width=1\linewidth]{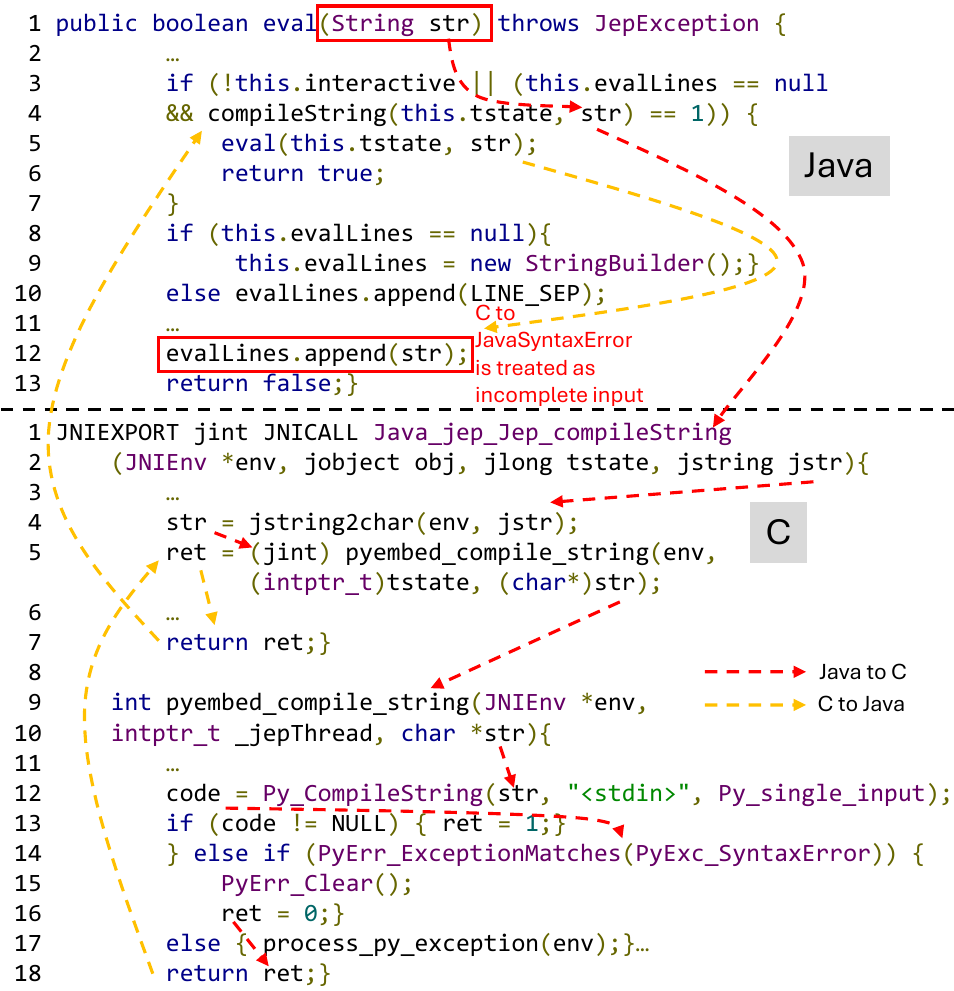}
    \caption{A vulnerability case found in Java-C project }
    \label{fig:jep}
\end{figure}
\vspace{-0pt}

{\tech} identified this vulnerability because it explicitly handles the \emph{macro-induced hidden connectivity}.
Concretely, the Python-visible constructor (e.g., \texttt{\small CMS\_Log8}) and the native initializer are connected through macro expansion rather than an explicit, name-consistent call/entry in the raw source. 
{\tech}'s macro handling (Macros Expanding to Function Calls) recovers these \emph{semantic} edges by resolving macro-expanded targets and injecting the corresponding CICFG edges, thereby restoring reachability from \tinycircnum{1} to the concrete C initializer.
Once this hidden edge is restored, {\tech}'s taint reasoning can propagate the user-controlled \texttt{\small width} across the language boundary to the allocation site (\tinycircnum{6}), and then flag the unsafe pattern that a potentially failing \texttt{\small calloc} result is stored into \texttt{\small self->table[i]} without a NULL check, enabling a later null-pointer dereference.

% The framework's explicit handling of cross-language interactions and its multi-stage analysis combining static techniques with LLM reasoning enables the detection of such vulnerabilities originating from parameter handling across language interfaces.

% \setminted{
%   highlightcolor=yellow
% }

% \begin{listing}[!ht]
% \scriptsize
% \begin{minted}[linenos]{python}
% if log_counting == 8:
%     self.cms = cmsc.CMS_Log8(width=self.width, depth=self.depth)
% \end{minted}
% \end{listing}
% \vspace{-20pt}
% \begin{listing}[!ht]
% \scriptsize
% \begin{minted}[linenos]{c}
% static int
% CMS_VARIANT(_init)(CMS_TYPE *self, PyObject *args, PyObject *kwds)
% {
%     static char *kwlist[] = {"width", "depth", NULL};

%     uint32_t w;
%     if (!PyArg_ParseTupleAndKeywords(args, kwds, "II", kwlist,
% 				      &w, &self->depth)) {
%         return -1;
%     }

%     if (self->depth  < 1 || self->depth > 32) {
%         char * msg = "Depth must be in the range 1-16";
%         PyErr_SetString(PyExc_ValueError, msg);
%     }

%     short int hash_length = -1;
%     while (0 != w)
%         hash_length++, w >>= 1;
%     if (hash_length < 0)
%         hash_length = 0;
%     self->width = 1 << hash_length;
%     self->hash_mask = self->width - 1;

%     HyperLogLog_init(&self->hll, 16);

%     self->table = (CMS_CELL_TYPE **) malloc(self->depth * sizeof(CMS_CELL_TYPE *));
%     int i;
%     for (i = 0; i < self->depth; i++)
%     {
%         self->table[i] = (CMS_CELL_TYPE *) calloc(self->width, sizeof(CMS_CELL_TYPE));
%     }
%     return 0;
% }
% \end{minted}
% \caption{Hello World in C}
% \label{listing:2}
% \end{listing}

For Java-C, we present a case where Java drives control flow and the native C layer returns a coarse-grained status that Java subsequently interprets, as shown in Figure~\ref{fig:jep}. In \texttt{eval(String str)} (Java L1--L6), the caller-provided \texttt{str} is deemed \emph{ready} only when \texttt{compileString(this.tstate, str) == 1}; otherwise, Java assumes the input is merely \emph{incomplete} and buffers it into the persistent \texttt{evalLines} state (Java L8--L13). The critical accumulation point is \texttt{evalLines.append(str)} (Java L12), which is executed whenever \texttt{compileString} does not return \texttt{1}.
This case also hinges on resolving \emph{polymorphism-induced dispatch} on the Java side: both \texttt{eval} and the \texttt{compileString(...)} invocation are ordinary instance-method calls, whose concrete targets are decided under Java's virtual dispatch semantics.
To avoid losing the reachability from \texttt{eval} to the native boundary, {\tech} applies Algorithm~\ref{algo:javaPolymorphism} to materialize the corresponding call/return edges for feasible targets, ensuring the Java control decision remains connected to the JNI entry.

The bug stems from a \emph{status-code semantic mismatch} across the Java--C boundary. On the native side, \texttt{Java\_jep\_Jep\_compileString} forwards \texttt{jstr} to \texttt{pyembed\_compile\_string} (C L1--L7, especially L4--L6). Inside \texttt{pyembed\_compile\_string}, multiple failure modes are collapsed into the same return value \texttt{ret = 0} (C L12--L17), notably after clearing an error and returning \texttt{0} for \texttt{SyntaxError} (C L14--L16). Since Java interprets \texttt{0} uniformly as ``not finished'', repeated invalid inputs are continuously appended rather than rejected, leading to unbounded buffer growth (DoS) at Java L12; if the same \texttt{Jep} instance is reused across trust boundaries, the persistent \texttt{evalLines} state can also enable cross-call state poisoning that is later consumed when the buffer is flushed. 
{\tech} detects this bug by combining feature-aware Java reachability recovery with cross-language reasoning: the \textbf{polymorphism feature handler} augments CICFG with virtual-dispatch call edges so that the Java control decision at \texttt{compileString(...)} remains transitively connected to the JNI implementation and its return code.

% Existing static, dynamic, and learning-based approaches likely failed to detect this specific vulnerability due to several factors. Static tools like MultiQL often rely on predefined query patterns and may struggle to precisely track data flow across Python-C API boundaries or reason about the semantic implications of insufficient input sanitization leading to allocation failures, particularly if specific rules for this pattern are absent. Dynamic analysis tools, such as PolyCruise or fuzzers like PolyFuzz, depend on runtime execution and would only trigger this bug under specific, hard-to-replicate conditions of memory exhaustion combined with a large input width. Achieving such conditions reliably during testing is challenging, and limited test coverage often hinders the discovery of such edge-case vulnerabilities. Learning-based tools like xLoc might not have been trained on data containing similar subtle memory management flaws stemming from cross-language parameter handling and missing null checks after allocation, limiting their ability to generalize and detect this specific pattern. {\tech}'s combination of cross-language control/data flow graph construction and LLM-assisted reasoning allows it to overcome these limitations by explicitly modeling the cross-language flow and reasoning about the potential consequences of the partially sanitized input.

\find{{\tech} discovered 17 previously unknown real-world bugs across 9 cross-language projects, including one assigned a CVE and another confirmed.  {\tech} recovers cross-language semantics that baseline analyses miss due to a lack of feature-aware reasoning.}

\section{Discussion}

\subsection{Why {\tech} works}

{\tech}’s success rests on three design principles that address the core challenges of static cross-language information flow analysis.

First, {\tech} builds comprehensive CICFGs. Conventional static data-flow analyzers often fail to stitch flows across language boundaries when complex features obscure control flow. In our motivating example~\ref{fig:vuln-example}, an analyzer may not treat the \texttt{\small @app.route} decorator as a source and may stop at an unresolved function pointer, preventing it from linking tainted Python input to the C format-string bug. {\tech} addresses this via a fixed-point iterative construction that incrementally enriches the CICFG until it captures the complete end-to-end control-flow path for reliable data-flow analysis.

Second, {\tech} uses an agentic LLM framework with the static analyzer as the controller. Instead of sending whole codebases to LLMs, the controller consults an LLM “expert panel” only for specific hard features that defeat traditional analysis, and mitigates hallucinations through multi-LLM negotiation and fact-checking.

Finally, {\tech} performs scalable two-phase refinement for data-flow reasoning. It first computes a fast taint over-approximation using control-flow reachability to isolate a small set of candidate paths, then applies LLM-driven taint refinement only to those paths, concentrating expensive semantic reasoning where it matters and keeping the analysis practical for real systems.

\subsection{Where {\tech} Falls Short and Why}
Despite its effectiveness, {\tech} makes trade-offs and depends heavily on its underlying components: the initial static analysis stack and the LLMs. Because it relies on Joern to generate the initial control-flow graphs, it inherits Joern’s parsing limitations and inaccuracies. More importantly, while LLMs provide the semantic reasoning to bridge language gaps, they are probabilistic and can be inaccurate, which in turn requires mitigation such as the Expert Panel.

{\tech}’s analytical depth also comes with performance costs. The same iterative feature handling and multi-stage, LLM-assisted taint refinement that enable discovery of subtle cross-language bugs make the analysis computationally intensive, leading to long runtimes on very large systems and limiting suitability for some development workflows. This is a deliberate rigor--speed trade-off.

Finally, real-world use and extension remain challenging. As with many static analyzers, {\tech} reports potential flows that may require manual validation of exploitability. Although modular by design, adding support for new language pairs is non-trivial and requires implementing specialized handlers and interaction rules for the new combination, demanding substantial domain expertise.

\section{Limitations of Existing Techniques}\label{sec:existlimits}

Static analysis frameworks such as MultiQL~\cite{youn2023declarative} struggle with semantic gaps in cross-language interactions. In our experiments, MultiQL detected no cross-language vulnerabilities in the benchmark projects, largely because it does not track data flows through Python object fields across the language boundary. This arises in common patterns: in {\small{\texttt{Aubio}}}, a C function is bound as a Python object method; MultiQL recognizes the object reference but cannot resolve the subsequent field access on the method call, so it misses the corresponding data-flow edge. As a result, it is ineffective for many Python–C systems.

Dynamic tools such as PolyCruise~\cite{wen22usenixsecurity} and PolyFuzz~\cite{wen23usenixsecurity} are limited by the execution paths they explore. Although they rediscovered the vulnerabilities they were designed to find, they found none of the new vulnerabilities uncovered by {\tech}, since effectiveness depends on test coverage from specific test cases or fuzzing campaigns. For example, the \texttt{\small bounter} vulnerability (Section~\ref{RQ4}) is triggered only when a large input causes memory allocation failure; dynamic approaches are unlikely to hit it unless inputs are crafted to induce memory exhaustion, which is difficult to trigger reliably.

Learning-based techniques such as xLoc~\cite{yang2024learning} also found no vulnerabilities in target projects, reflecting their inability to generalize beyond bug patterns represented in training data. For the \texttt{\small bounter} case, a model would need training examples of cross-language parameter handling that leads to unchecked allocation failures; without such examples, the pattern is unlikely to be recognized.

% {\tech} was designed to overcome these specific limitations by combining the comprehensive path coverage of static analysis with the semantic reasoning required to navigate complex cross-language interactions.
% root: paper-ccs.tex

\section{Related Work}\label{sec:related}
%\vspace{3pt}\noindent
\textbf{LLM-assisted/agent-based program analysis.}
Given their capabilities for code understanding and generating, LLMs have been used to assist with various program analysis tasks, ranging from comprehending source/obfuscated code~\cite{fang2024large} to detecting/classifying software vulnerabilities~\cite{linlarge25ndss}. 
{LLift}~\cite{li2024enhancing} complements symbolic execution with LLMs for identifying use-before-initialization bugs in the Linux kernel, utilizing the LLMs to extract and solve path constraints that trigger the bugs. 
To recover names and types for local variables and user-defined data structures from stripped binaries, {ReSym}~\cite{xie2024resym} fine-tunes two LLMs and combines them with Prolog-based reasoning, mimicking a manual reverse engineering process. 
Recently, {IRIS}~\cite{li2025iris} leverages LLMs to assist CodeQL with static taint analysis for Java projects, using LLMs to refine CodeQL's taint specification and then to triage the taint paths reported by CodeQL via contextual analysis. 
Li \etal~\cite{li2025fine} construct a cross-language bug dataset spanning Python--C, Java-C, and Python-Java interaction mechanisms, and fine-tune CodeLMs to predict buggy vs.\ clean cross-language functions.
% However, it remains a snippet/function-level predictor and does not reconstruct interprocedural cross-language control/data-flow evidence needed for precise flow reasoning.
AutoCodeRover~\cite{zhang2024autocoderover} builds an agent by combining LLMs with code search to automatically solve GitHub issues, while 
SWE-Agent~\cite{yang2024swe} and OpenHands~\cite{wang2024openhands} are agents serving a broader set of coding tasks such as code editing. 

% In comparison, {\tech} addresses multilingual code while involving LLMs directly in the core program analysis steps (control-flow augmentation and data-flow propagation) and without fine-tuning. Our agentic approach is also different in that it adds autonomy to LLMs via program analysis, rather than merely letting LLMs assist with the latter. 

\vspace{3pt}\noindent
\textbf{Static taint/information flow analysis.}
Language-level information flow analysis, such as JFlow~\cite{myers1999jflow} for Java, realizes static flow checking by extending a given language, while application-level 
approaches like Pixy~\cite{jovanovic2006pixy} for PhP detect taint-style vulnerabilities via data-flow analysis. 
Static taint analysis has also been enabled to detect sensitive user inputs in SUPOR~\cite{huang2015supor} and privacy leaks in UIPicker~\cite{nan2015uipicker}, both 
for Android apps and leveraging FlowDroid~\cite{arzt2014flowdroid}. 
Information flow control has also been achieved at the level of the hardware description language to verify hardware security mechanisms~\cite{ferraiuolo2017verification}. 
In contrast, we focus on static information flow analysis \textit{across different languages}, versus these approaches addressing flows within language boundaries. 

\vspace{3pt}\noindent
\textbf{Cross-language code analysis.}
By adopting CodeQL to multi-language projects, MultiQL~\cite{youn2023declarative} achieves proof-of-concept static cross-language bug detection by extending CodeQL's 
databases to include interoperation rules based on cross-language APIs. 
% The reliance on CodeQL makes MultiQL inherit its fundamental limitations, such as applicability to only vulnerabilities whose code patterns are known and can be well-defined, as well as low precision and recall due to the lack of in-depth data/control flow analysis and ignorance of challenging language features. 
PowerPoly~\cite{jiang2025powerpoly} translates multilingual programs into a unified WebAssembly-based IR to eliminate language boundaries, and performs vulnerability detection via static analysis plus dynamic analyses.
However, it relies on faithful source-to-Wasm translation, and the low-level Wasm IR may obscure language-specific semantics.
% , potentially limiting precise source-level interprocedural reasoning.
AXA~\cite{roth2024axa} obtains cross-language static analysis results by orchestrating the interleaved execution of single-language analyses, propagating results between
the two analyzers hence integrating their results. 
Similarly, JNIFER~\cite{zhang2025interactive} interactively resolves native code to accomplish cross-language pointer analysis, while 
CSS~\cite{kan2024cross} refines the native code specification by incorporating caller sensitivity, both for Java-C code. 
A number of other static analysis approaches for multilingual programs exist, relying on handcrafted summaries/rules an/or heavy engineering, as summarized in~\cite{roth2024axa}. 
% However, like JNIFER and CSS, AXA relies on the availability and capabilities of a sophisticated analysis infrastructure of each language. 
Orthogonally, CrossLangFuzzer~\cite{feng2025finding} generates cross-language test programs for JVM-based languages and applies differential testing to expose compiler bugs across multiple JVM-language compilers.
% This line of work is complementary but targets compiler correctness, rather than cross-language security bugs in multilingual applications.

% {\tech} is developed to overcome these limitations of existing approaches, explicitly handling various language features by leveraging the merits of LLMs for code analysis.  

% root: paper-ccs25.tex
\section{Conclusion}\label{sec:conclusion}
With cross-language vulnerabilities burgeoning and continuing to compromise the security of multi-language systems, 
it is imperative to discover those defects before they are exploited. 
Yet existing defensive techniques suffer from various critical limitations that impede their practical adoption. 
We thus explore a neuro-symbolic approach to statically reasoning about information flow across language boundaries, hence 
developing {\tech}, a static cross-language information flow analysis framework. 
By combining the strengths of traditional static analysis and large-language-models (LLMs) while mitigating their respective weaknesses, 
{\tech} overcomes semantics disparities and challenging language features within and across heterogeneous languages.  
%explicitly dealing with key technical challenges to this analysis 
Compared to state-of-the-art solutions of various kinds against diverse real-world multi-language systems, 
{\tech} demonstrated its merits in both cost-effectiveness and cross-language vulnerability discovery. 

{\tech} has been implemented to work with two impactful language combinations, Python-C and Java-C, which dominates 
machine learning and web/mobile software ecosystems, respectively, and remains extensible to support other language combinations.

{
% \footnotesize 
 %\small
 %\bibliographystyle{unsrt}
 %\bibliographystyle{plain}
\bibliographystyle{plain}
 % \bibliographystyle{ieeetr}
 %\bibliography{reference-abbr}
 \bibliography{paper-cai}
}

\appendices

\label{lab:appendix}

\section{C Feature Handling Algorithms}\label{sec:featurehandlingalgos-c}
This section presents the representative C feature handler\,---\,Conditional Compilation\,---\,that augments a CICFG with additional interprocedural edges for a language construct whose execution targets are indirect, deferred, or runtime-resolved. The pseudocode algorithms for the remaining six C feature handlers (Function Pointers, Dynamic Linking, Inline Assembly, Macros to Function Calls, Goto Statements, and Setjmp/Longjmp) follow the same template and are provided in our artifact package.

\subsection{Conditional Compilation }\label{sec:c-feature-condcomp}

\begin{algorithm}[!htbp]
\scriptsize
\caption{\footnotesize{Conditional Compilation Analysis (C feature)}}
\label{algo:c-condcomp}
\SetKwProg{Fn}{Function}{}{end}
\SetKwFunction{AnalyzeCondComp}{AnalyzeCondComp}
\SetKwFunction{PanelQuery}{PanelQuery}
\SetKwFunction{UpdateCICFGNode}{UpdateCICFGNode}
\SetKwFunction{UpdateCICFGEdge}{UpdateCICFGEdge}
\SetKw{Return}{return}
\SetKw{Continue}{continue}
\LinesNumbered
\KwIn{Set of files $F$}
\KwOut{Updated CICFG (nodes/edges for alternative preprocessor branches)}

\Fn{\AnalyzeCondComp{$F$}}{
    touched $\gets$ []; is\_updated $\gets$ false\;
    \ForEach{$file \in F$}{\label{cc:iterF}
        res $\gets$ \PanelQuery{"Does this file contain conditional compilation directives? Output YES/NO only." + fileContent(file)}\;\label{cc:check}
        \If{res == "NO"}{\Continue}
        touched.append(file)\;
        stmts, edges $\gets$ \PanelQuery{
        "Enumerate statements guarded by preprocessor branches and the feasible control-flow edges across branches.
         Output stmts as $<(line, stmtText)>$ and edges as $<(srcLine, dstLine, tag)>$." + fileContent(file)}\;\label{cc:extract}
        \ForEach{$(line, s) \in stmts$}{\label{cc:addNode}
            \UpdateCICFGNode{file, line, s, \{"feat":"condcomp"\}}\;
        }
        \ForEach{$(u, v, tag) \in edges$}{\label{cc:addEdge}
            \UpdateCICFGEdge{file, u, file, v, \{"feat":"condcomp","branch":tag\}}\;
            is\_updated $\gets$ true\;
        }
    }
    \Return{touched, is\_updated}\;
}
\end{algorithm}

\paragraph{Description.}
Conditional compilation materializes \emph{configuration-dependent} control-flow that a single preprocessed build cannot cover: each \texttt{\#ifdef}, \texttt{\#ifndef}, \texttt{\#if}, \texttt{\#else}, \texttt{\#endif} family selects a mutually exclusive code region, and statements that are syntactically present in the source may be excised\,---\,or, conversely, only become reachable\,---\,under a particular build configuration. A standard CICFG built from a single preprocessed view therefore systematically loses statements and edges contributed by alternative branches, which biases every downstream reachability/taint computation.

Algorithm~\ref{algo:c-condcomp} repairs this in three steps. \textbf{(1)~Detection (line~\ref{cc:check}).} For every input file, the expert panel is asked a YES/NO question on whether preprocessor branching is present; files without any directive are skipped to avoid spending LLM budget on regular C code. \textbf{(2)~Branch enumeration (line~\ref{cc:extract}).} On a positive file, the panel returns two structured lists: \emph{stmts}\,---\,statements that exist only inside a particular branch, each tagged with its source line and text\,---\,and \emph{edges}\,---\,feasible control-flow edges between lines, each tagged with the branch it belongs to. The panel produces this view jointly across all branches, including ones that the active build configuration would have hidden. \textbf{(3)~CICFG augmentation (lines~\ref{cc:addNode}--\ref{cc:addEdge}).} Each branch-guarded statement is materialized as a CICFG node tagged \texttt{"feat":"condcomp"}, and each enumerated edge is materialized between the corresponding source/destination lines, additionally tagged with its branch identifier so that later analyses can reason \emph{per configuration} (intra-branch edges) or \emph{across configurations} (inter-branch edges) on the same graph. The handler returns the list of touched files and a boolean signaling whether new edges were introduced, which feeds back into the controller's worklist.

\paragraph{Illustrating example.}
\begin{lstlisting}[language=C, firstnumber=1, caption={Conditional compilation yields multiple mutually-exclusive paths.}, label={lst:c-condcomp}]
#ifdef _MSC_VER          // L1
#include <intrin.h>      // L2  (only under MSVC)
#endif                   // L3
#include <arm_neon.h>    // L4  (always included)

int main(int argc, char **argv) {        // L6
  /* ... */
#ifdef __aarch64__                       // L8
  /* AArch64-specific fast path */
  ret += (int)vgetq_lane_f64(vfmaq_f64(vd1, vd2, vd3), 0);  // L10
#endif                                   // L11
  return ret;                            // L12
}
\end{lstlisting}
On this example, the detection step (line~\ref{cc:check}) returns YES because two \texttt{\#ifdef} directives are present (\texttt{\_MSC\_VER} and \texttt{\_\_aarch64\_\_}). The branch-enumeration step (line~\ref{cc:extract}) yields two branch-guarded statements\,---\,the \texttt{<intrin.h>} include at L2 (gated by \texttt{\_MSC\_VER}) and the AArch64 NEON computation at L10 (gated by \texttt{\_\_aarch64\_\_})\,---\,plus the corresponding intra-branch edges (e.g., L8$\to$L10$\to$L11 inside the AArch64 branch) and the inter-branch fall-through L11$\to$L12 that re-joins the unconditional control flow. The augmentation step inserts L2 and L10 as new \texttt{condcomp} nodes (so neither statement is silently dropped under the ``other'' configuration) and adds the listed edges with their branch tags, so a downstream taint analysis traversing \texttt{main} can reach the NEON intrinsic regardless of which configuration the developer's local build happens to select.

% ===================================================================
%                    Python features
% ===================================================================

\section{Python Feature Handling Algorithms}
\label{sec:featurehandlingalgos-python}

This section presents the representative Python feature handler\,---\,First-Class Functions\,---\,that augments a CICFG with additional interprocedural edges for a language construct whose execution targets are indirect, deferred, or runtime-resolved. The pseudocode algorithms for the remaining five Python feature handlers (Dynamic Typing, Decorators, Lambda Expressions, Reflection, and Dynamic Imports) follow the same template and are provided in our artifact package.

\setlength{\textfloatsep}{0pt}

% ===================================================================
% Python feature: first-class functions
\subsection{First-class functions}

\begin{algorithm}[!htbp]
\scriptsize
\caption{\footnotesize{First-Class Functions Analysis - Python feature}}
\label{algo:firstClassFunctionsAnalysis}
\SetKwProg{Fn}{Function}{}{end}
\SetKwFunction{AnalyzeFirstClassFunctions}{AnalyzeFirstClassFunctions}
\SetKwFunction{ResolveHigherOrderFlow}{ResolveHigherOrderFlow}
\SetKwFunction{SendToLLM}{SendToLLM}
\SetKwFunction{UpdateCICFGEdge}{UpdateCICFGEdge}
\SetKw{Continue}{continue}
\SetKw{Return}{return}
\LinesNumbered
\KwIn{Set of files $F$ from a Python program, CICFG}
\KwOut{Updated CICFG}

\Fn{\AnalyzeFirstClassFunctions{F}}{
    start\_file $\gets$ [\,]\;
    is\_updated $\gets$ false\;

    \ForEach{file in $F$}{
        flag $\gets$ \SendToLLM{
        "Given the following Python file, determine whether it uses first-class functions (functions passed as values, e.g., as arguments or returned). "
        "Output 'YES' or 'NO' only. No explanation. "
        + \textit{file content}}\;

        \If{flag == "NO"}{
            \Continue\;
        }
        start\_file.append(file)\;

        resp $\gets$ \SendToLLM{
        "List all sites in this file where a user-defined function is passed as an argument to another user-defined function. "
        "For each site output a tuple <passed\_func, receiver\_func, receiver\_param, call\_line>. "
        "Here receiver\_param is the formal parameter of receiver\_func that receives passed\_func at that call site. "
        "If none, output 'NO'. No explanation. "
        + \textit{file content}}\;

        \If{resp == "NO"}{
            \Continue\;
        }

        % passList $\gets$ parse resp into list of (passed\_func, receiver\_func, receiver\_param, call\_line)
        \ForEach{(passed\_func, receiver\_func, receiver\_param, call\_line) in resp}{
            visited $\gets$ set()\;
            is\_updated $\gets$ \ResolveHigherOrderFlow{passed\_func, receiver\_func, receiver\_param, visited} \textnormal{\textbf{or}} is\_updated\;
        }
    }

    \Return{start\_file, is\_updated}\;
}

\Fn{\ResolveHigherOrderFlow{passed\_func, receiver\_func, receiver\_param, visited}}{
    \If{(passed\_func, receiver\_func, receiver\_param) in visited}{
        \Return{false}\;
    }
    visited.add((passed\_func, receiver\_func, receiver\_param))\;

    receiver\_file, receiver\_entry $\gets$ retrieve the file name and entry line number of \textit{receiver\_func} from CICFG\;
    code $\gets$ retrieve the body of \textit{receiver\_func} from CICFG\;

    res $\gets$ \SendToLLM{
    "Within the following function body, analyze the parameter '" + receiver\_param + "'. "
    "Output two lists: "
    "(1) call\_lines = all line numbers where '" + receiver\_param + "' is invoked as a callable (e.g., " + receiver\_param + "(...)); "
    "(2) forward\_sites = all sites where '" + receiver\_param + "' is forwarded as an argument to some function g; output each as <g, g\_param> "
    "where g\_param is the formal parameter of g receiving it. "
    "If neither exists, output 'NO'. No explanation. "
    + \textit{function body}}\;

    \If{res == "NO"}{
        \Return{false}\;
    }

    updated $\gets$ false\;

    \If{res.call\_lines is not empty}{
        target\_file, target\_entry $\gets$ retrieve the file name and entry line number of \textit{passed\_func} from CICFG\;
        \ForEach{ln in res.call\_lines}{
            annotation $\gets$ \{"first-class": receiver\_param + "->" + passed\_func\}\;
            \UpdateCICFGEdge{receiver\_file, ln, target\_file, target\_entry, annotation}\;
            updated $\gets$ true\;
        }
    }

    \If{res.forward\_sites is not empty}{
        \ForEach{(g, g\_param) in res.forward\_sites}{
            updated $\gets$ \ResolveHigherOrderFlow{passed\_func, g, g\_param, visited} \textnormal{\textbf{or}} updated\;
        }
    }

    \Return{updated}\;
}
\end{algorithm}

\paragraph{Description.}
First-class functions enable higher-order control flow: a function value is passed into another function and invoked indirectly via a parameter, so the callee name is not syntactically present at the invocation site and a call graph that only resolves direct calls silently drops the entire higher-order edge.

Algorithm~\ref{algo:firstClassFunctionsAnalysis} restores these edges in two stages. \textbf{(1)~Per-file extraction (\texttt{AnalyzeFirstClassFunctions}).} The expert panel first answers a YES/NO probe on whether the file uses functions-as-values; on a positive answer, it returns every concrete \emph{passing site} as a tuple $\langle \mathit{passed\_func}, \mathit{receiver\_func},$ $\mathit{receiver\_param}, \mathit{call\_line}\rangle$, where \texttt{passed\_func} is the user-defined function being passed as a value, \texttt{receiver\_func} is the higher-order function it is passed into, and \texttt{receiver\_param} is the formal parameter inside \texttt{receiver\_func} that will receive it. \textbf{(2)~Higher-order resolution (\textit{ResolveHigherOrderFlow}).} For each passing site, the analysis enters the body of \texttt{receiver\_func} and asks the panel for two things: \emph{call\_lines}\,---\,every line at which \texttt{receiver\_param} is invoked as a callable\,---\,and \emph{forward\_sites}\,---\,every line at which \texttt{receiver\_param} is forwarded onward as an argument to a further function $g$ via some formal parameter \texttt{g\_param}. Each call line yields a new CICFG edge from that line to the entry of \texttt{passed\_func}, annotated as \texttt{first-class}; each forward site causes the analysis to recurse into $g$ with $\langle\mathit{passed\_func}, g, g\_param\rangle$, so a value that traverses several higher-order layers is still chased to its eventual invocation. A \texttt{visited} set guards against revisits in the presence sof mutual or cyclic forwarding, and the routine returns whether any new edge was added so the controller knows whether to re-run dependent feature handlers.

\paragraph{Illustrating example.}
\begin{lstlisting}[language=Python, firstnumber=1, caption={Higher-order function: passing a function as a value (simplified).}, label=fig:first-class]
def my_map(func, arg_list):     # L1  receiver_func = my_map; receiver_param = func
    result = []                 # L2
    for i in arg_list:          # L3
        result.append(func(i))  # L4  invocation of receiver_param  -> call_line
    return result               # L5

def square(x):                  # L7  passed_func = square
    return x * x                # L8

# passing square without '()' does not execute it; it can be executed later inside my_map()
squares = my_map(square, [1, 2, 3, 4, 5])   # L11  passing site
\end{lstlisting}
On this example, the per-file extraction yields a single passing-site tuple $\langle\texttt{square}, \texttt{my\_map}, \texttt{func}, \mathrm{L11}\rangle$. The higher-order resolution then enters \texttt{my\_map}'s body and the panel identifies $\texttt{call\_lines}{=}\{\mathrm{L4}\}$ (because \texttt{func} is invoked as \texttt{func(i)} at L4) and $\texttt{forward\_sites}{=}\emptyset$ (\texttt{func} is not passed onward). The handler therefore inserts a single CICFG edge from L4 of \texttt{my\_map} to the entry of \texttt{square}, annotated \texttt{\{"first-class": "func->square"\}}, making the higher-order call \texttt{func(i)}$\,\to\,$\texttt{square} explicit for downstream reachability and taint analyses. Had \texttt{square} been forwarded through one or more wrapper layers before reaching its invocation, the recursive \texttt{ResolveHigher- OrderFlow} call would have followed it through each layer until a \texttt{call\_line} was found.

% ======================================================================
% Reviewed and expanded on 2026-02-05 (Asia/Taipei).
% This section is intended for direct inclusion in appendix_all.tex.
% ======================================================================

\section{Java Feature Handling Algorithms}
\label{sec:featurehandlingalgos-java}

\noindent
This section presents the representative Java feature handler\,---\,Polymorphism (i.e., polymorphic virtual dispatch on member methods)\,---\,that augments a CICFG with additional interprocedural edges for a language construct whose execution targets are indirect, deferred, or runtime-resolved. The pseudocode algorithms for the remaining four Java feature handlers (Reflection, Lambda Expressions, Dynamic Proxies, and Dynamic Class Loading) follow the same template and are provided in our artifact package.
% Each handler inserts paired \emph{call} and \emph{return} edges between statement-level nodes identified by (file name, line number).
% The procedures use an LLM-assisted extractor (\texttt{SendToLLM}) as a semantic oracle for localizing feature-specific events (e.g., reflective invocations, lambda bodies, proxy callsites) and for producing structured records used by the CICFG augmentation pass.

% (Java-specific summary table removed; now consolidated into the master table in Appendix~\ref{sec:feature_handler_summary}.)

% ----------------------------------------------------------------------
% Java feature 5
% ----------------------------------------------------------------------

\subsection{Polymorphism}\label{sec:java-feature-polymorphism}

\setlength{\textfloatsep}{0pt}
%\begin{algorithm*}[tp]
\begin{algorithm*}[!htbp]
\scriptsize
\caption{\footnotesize{Polymorphism Analysis -- Java feature}}
\label{algo:javaPolymorphism}
\SetKwProg{Fn}{Function}{}{end}
\SetKwFunction{AnalyzePolymorphism}{AnalyzePolymorphism}
\SetKwFunction{DetectVirtualCalls}{DetectVirtualCalls}
\SetKwFunction{ExtractVirtualCalls}{ExtractVirtualCalls}
\SetKwFunction{ResolveMethodSig}{ResolveMethodSig}
\SetKwFunction{GetDeclType}{GetDeclType}
\SetKwFunction{CHA}{CHA}
\SetKwFunction{PointsToTypes}{PointsToTypes}
\SetKwFunction{RefineTypesLLM}{RefineTypesLLM}
\SetKwFunction{Overrides}{Overrides}
\SetKwFunction{ResolveImpl}{ResolveImpl}
\SetKwFunction{FilenameFromCICFG}{FilenameFromCICFG}
\SetKwFunction{EntryLine}{EntryLine}
\SetKwFunction{ReturnLine}{ReturnLine}
\SetKwFunction{UpdateCICFGEdge}{UpdateCICFGEdge}
\SetKwFunction{RetrieveFunctionBody}{RetrieveFunctionBody}
\SetKwFunction{SendToLLM}{SendToLLM}
\SetKw{Return}{return}
\LinesNumbered
\KwIn{set of Java methods $F$; class table $C$; (optional) points-to map $\mathcal{PT}$; CICFG keyed by (file, line)}
\KwOut{(start\_methods, is\_updated) and CICFG augmented with polymorphism-induced call/return edges}

\Fn{\AnalyzePolymorphism{$F, C, \mathcal{PT}$}}{
    start\_methods $\gets$ []\;
    is\_updated $\gets$ false\;

    \ForEach{$m \in F$}{
        code $\gets$ \RetrieveFunctionBody{$m$}\;
        \If{\DetectVirtualCalls{code} == false}{continue}

        start\_methods.append($m$)\;

        calls $\gets$ \ExtractVirtualCalls{code}\;
        \ForEach{c $\in$ calls}{
            \tcp{c schema: \{call\_line, recv\_expr, recv\_var, static\_recv\_type, method\_name, param\_types, ret\_type\}}
            caller\_file $\gets$ \FilenameFromCICFG{$m$}\;
            sig $\gets$ \ResolveMethodSig{c.method\_name, c.param\_types, c.ret\_type}\;

            \tcp{Step 1: obtain candidate dynamic receiver types}
            T0 $\gets$ c.static\_recv\_type\;
            Cand $\gets$ \{\}\;

            \If{$\mathcal{PT}$ is available}{
                Cand $\gets$ \PointsToTypes{$\mathcal{PT}$, $m$, c.recv\_var}\;
            }
            \If{Cand is empty}{
                \tcp{Conservative CHA fallback: all subtypes of the static receiver type}
                Cand $\gets$ \CHA{$C$, T0}\;
            }

            \tcp{Optional refinement: use local allocation/guards to prune Cand}
            Cand' $\gets$ \RefineTypesLLM{code, c, Cand}\;
            \If{Cand' not empty}{Cand $\gets$ Cand'}\;

            \tcp{Step 2: resolve overriding target methods for each candidate type}
            Targets $\gets$ \{\}\;
            \ForEach{$T \in$ Cand}{
                \tcp{Resolve the concrete implementation: prefer override, else inherit}
                t $\gets$ \ResolveImpl{$C$, $T$, sig}\;
                \If{$t \neq$ NULL}{Targets.add($t$)}
            }

            \tcp{Step 3: add CICFG call/return edges for all possible targets}
            \ForEach{$t \in$ Targets}{
                callee\_file $\gets$ \FilenameFromCICFG{$t$}\;
                entry $\gets$ \EntryLine{$t$}\;
                ret $\gets$ \ReturnLine{$t$}\;

                \UpdateCICFGEdge{caller\_file, c.call\_line, callee\_file, entry,
                    \{\"poly\":\"virtual\", \"recv\":T0, \"sig\":sig\}}\;
                \UpdateCICFGEdge{callee\_file, ret, caller\_file, c.call\_line,
                    \{\"poly\":\"virtual\", \"recv\":T0, \"sig\":sig\}}\;
                is\_updated $\gets$ true\;
            }
        }
    }

    \Return{start\_methods, is\_updated}\;
}

\Fn{\DetectVirtualCalls{code}}{
    ans $\gets$ \SendToLLM{detect polymorphic dispatch sites (interface/base-typed receiver call); return JSON \{result\}}\;
    \Return{(ans.result == "YES")}\;
}

\Fn{\ExtractVirtualCalls{code}}{
    ans $\gets$ \SendToLLM{extract virtual call records with receiver/type/signature and call lines; return JSON \{result, calls[]\}}\;
    \eIf{ans.result == "YES"}{\Return{ans.calls}}{\Return{$[\ ]$}}
}

\Fn{\RefineTypesLLM{code, c, Cand}}{
    ans $\gets$ \SendToLLM{prune Cand using local evidence (alloc sites, instanceof checks, guards); return \{result, pruned\_types[]\}}\;
    \eIf{ans.result == "YES"}{\Return{set(ans.pruned\_types)}}{\Return{$[\ ]$}}
}

\Fn{\ResolveImpl{$C$, $T$, sig}}{
    \tcp{Return the most specific implementation of sig for dynamic type T if exists}
    \If{\Overrides{$C$, $T$, $sig$} $!=$ NULL}{\Return{\Overrides{$C$, $T$, $sig$}}}
    \tcp{Otherwise, walk supertypes until a declaration is found}
    \Return{firstDeclInSuperChain($C$, $T$, $sig$)}\;
}
%\end{algorithm*}
\end{algorithm*}

\paragraph{Description.}
Java polymorphic method calls on an interface-typed or base-class-typed receiver (i.e., \texttt{invokeinterface}/\texttt{invokevirtual} bytecodes) are resolved at runtime against the receiver's \emph{dynamic} type, not its declared static type. A purely syntactic call graph that links a callsite only to the method declared on the static type therefore systematically misses every concrete override, breaking cross-procedural reachability and dropping data-flow paths that traverse the dispatched body.

Algorithm~\ref{algo:javaPolymorphism} reconstructs these dispatched edges in three steps. \textbf{(1)~Virtual-callsite extraction ({DetectVirtualCalls} / {ExtractVirtualCalls}).} For every method $m\in F$, the expert panel first answers a YES/NO probe on whether $m$ contains a polymorphic dispatch site, and on a positive answer it returns one record per callsite, schema 
%$\langle 
<\texttt{call\_line}, \texttt{recv\_expr}, \texttt{recv\_var}, \texttt{static\_recv\_type}, \texttt{method\_name}, \texttt{param\_types}, \texttt{ret\_type}>. 
%\rangle$. 
The signature \texttt{sig} is then derived from \texttt{(method\_name, param\_types, ret\_type)} so it can be looked up uniformly inside the class table $C$. \textbf{(2)~Candidate dynamic-type set.} The handler tries to bound the dynamic type of \texttt{recv\_var} as tightly as soundness allows: if a points-to map $\mathcal{PT}$ is available, it queries \texttt{PointsToTypes($\mathcal{PT}, m, \texttt{recv\_var}$)}; otherwise it falls back to Class Hierarchy Analysis, taking every subtype of \texttt{static\_recv\_type} from $C$ as a candidate. The candidate set may then be pruned by \texttt{RefineTypesLLM}, which inspects local evidence (e.g., \texttt{new T(...)} allocations and \texttt{instanceof} guards) to drop infeasible types, but only when this strictly shrinks the set\,---\,never enlarges it. \textbf{(3)~Per-target edge insertion (\texttt{ResolveImpl} + \texttt{UpdateCICFGEdge}).} For each candidate type $T$, \texttt{ResolveImpl} walks the override chain (\texttt{Overrides($C,T,\texttt{sig}$)}, falling back to the first declaration in $T$'s super-chain) to identify the concrete implementation that runtime virtual dispatch would actually pick; the handler then inserts a paired \emph{call} edge (callsite line $\to$ callee entry) and \emph{return} edge (callee return $\to$ callsite line), both annotated \texttt{\{"poly":"virtual","recv":T0,"sig":sig\}}. The \texttt{is\_updated} flag and \texttt{start\_methods} list propagate to the controller so that this handler\,---\,which is propagation-marked (Y)\,---\,can be re-run after dependent feature handlers (e.g., reflective allocation) widen the candidate set.

\paragraph{Illustrating example.}
\begin{lstlisting}[language=Java, firstnumber=1, caption=Polymorphism example aligned with Algorithm~\ref{algo:javaPolymorphism}., label=fig:java_poly_example]
package demo;

interface Sink { void write(String s); }                 // L3

final class FileSink implements Sink {
  public void write(String s) { System.out.println("F:"+s); } // L6
}

final class NetSink implements Sink {
  public void write(String s) { System.out.println("N:"+s); } // L10
}

public final class PolyDemo {
  static Sink choose(boolean flag) {                     // L14
    return flag ? new FileSink() : new NetSink();        // L15
  }

  static void run(boolean flag, String data) {           // L18
    Sink sink = choose(flag);                            // L19  (sink: static type Sink)
    sink.write(data);                                    // L20  (virtual dispatch callsite)
  }
}
\end{lstlisting}
On this example, step~(1) returns one virtual-callsite record for \texttt{run}, with \texttt{call\_line=L20}, \texttt{recv\_var=sink}, \texttt{static\_recv\_type=Sink}, and signature \texttt{write(String):void}. Step~(2) yields the candidate set $\{\texttt{FileSink},\texttt{NetSink}\}$\,---\,either because points-to evidence flows the two allocations on L15 forward through \texttt{choose}'s return into \texttt{sink}, or, in the absence of points-to information, because CHA enumerates every subtype of \texttt{Sink}. The optional \texttt{RefineTypesLLM} prune leaves the set unchanged here (no \texttt{instanceof} guard rules either type out). Step~(3) then resolves each candidate to its concrete override\,---\,\texttt{FileSink.write(String)} (entry/return at L6) and \texttt{NetSink. write(String)} (entry/return at L10)\,---\,and inserts two paired call/return edges from L20 to each entry/return, both annotated \texttt{\{"poly": "virtual", }\allowbreak\texttt{"recv": "Sink", }\allowbreak\texttt{"sig": "write(String):void"\}}. Downstream taint analysis traversing \texttt{run} can now reach \emph{both} \texttt{println} sites at L6 and L10 from \texttt{data}, instead of stopping at the unresolved \texttt{Sink.write} declaration on L3.

\end{document}